\documentclass[reqno,11pt]{amsart}
\usepackage{amsmath,amssymb,amsthm,graphicx,a4wide,enumerate,url}
\usepackage[small,bf]{caption} 
\usepackage{subfig}
\begin{document}
\parskip.9ex

%===========================================================================
%=================================================================== Titles.
\title[Traffic Control and Fuel Consumption Reduction via Moving Bottlenecks]
{Traffic Flow Control and Fuel Consumption Reduction via Moving Bottlenecks}
\author[R. A. Ramadan]{Rabie A. Ramadan}
\address[Rabie A. Ramadan]
{Department of Mathematics \\ Temple University \\ \newline
1805 North Broad Street \\ Philadelphia, PA 19122}
\email{rabie.ramadan@temple.edu}
\author[B. Seibold]{Benjamin Seibold}
\address[Benjamin Seibold]
{Department of Mathematics \\ Temple University \\ \newline
1805 North Broad Street \\ Philadelphia, PA 19122}
\email{seibold@temple.edu}
\urladdr{http://www.math.temple.edu/\~{}seibold}

%===========================================================================
\begin{abstract}
Moving bottlenecks, such as slow-driving vehicles, are commonly thought of as impediments to efficient traffic flow. Here, we demonstrate that in certain situations, moving bottlenecks---properly controlled---can actually be beneficial for the traffic flow, in that they reduce the overall fuel consumption, without imposing any delays on the other vehicles. As an important practical example, we study a fixed bottleneck (e.g., an accident) that has occurred further downstream. This new possibility of traffic control is particularly attractive with autonomous vehicles, which (a) will have fast access to non-local information, such as incidents and congestion downstream; and (b) can execute driving protocols accurately.
\end{abstract}
%===========================================================================

\maketitle

%===========================================================================
\section{Introduction}
%===========================================================================
This paper demonstrates that, in certain situations, traffic flow can be controlled via a single moving bottleneck so that the overall fuel consumption is reduced; yet, no delay is imposed on the travel time of the other vehicles. An important scenario is the control of traffic flow upstream of a fixed bottleneck, where traffic is still in free-flow.

A fixed bottleneck (FB) is a region on a road (we focus solely on highways here) at which the throughput is reduced. Common examples are work zones, road features (curves, climbs), and traffic incidents. Throughput reduction can be caused by lane reduction, speed reduction, or both. A moving bottleneck (MB) follows the same principles, however, it moves along the road. Common examples are slow-moving vehicles or moving road work zones (see \cite{Daganzo1997} for the fundamentals and traffic flow theory of bottlenecks). Here we focus on bottlenecks with regions of influence much shorter than the length scales of interest, so we model them as fixed or moving points (with zero length).

The evolution of the traffic density $\rho(x,t)$ along the road is modeled macroscopically via the Lighthill-Whitham-Richards (LWR) model \cite{LighthillWhitham1955, Richards1956}
\begin{equation}
\label{eq:LWR}
\rho_t+(Q(\rho))_x = 0\;,
\end{equation}
where $Q(\rho) = \rho U(\rho)$ is the flux function, encoding the fundamental diagram (FD), and $U(\rho)$ is the bulk velocity vs.~density relationship. Being a hyperbolic conservation law, Eq.~\eqref{eq:LWR} models sharp transition zones (e.g., upstream ends on traffic jams) as moving discontinuities (``shocks''). A shock between two states $\rho_1<\rho_2$ moves at a speed $s = (Q(\rho_2)-Q(\rho_1))/(\rho_2-\rho_1)$, given by the slope of the secant line in the FD connecting the two states.

While the techniques presented herein apply for general FD shapes, we consider two specific examples: the quadratic Greenshields \cite{Greenshields1935} flux $Q(\rho) = \rho\,u_\text{m}(1-\rho/\rho_\text{m})$, and the triangular Newell--Daganzo \cite{Newell1993, Daganzo1994} flux. The former is used to demonstrate the theory for a strictly concave flux; and the latter---better resembling true data \cite{Greenberg1959}---is used to obtain quantitative estimates.

Regarding the model description, it should be stressed that the LWR model captures---even with the best FD---only the large scale equilibrium/bulk flow. On a microscopic scale, vehicles may exhibit non-equilibrium behavior (such as velocity oscillations), particularly near bottlenecks. Such effects can in principle be described by second-order macroscopic models \cite{Whitham1974, Payne1979, Lebacque1993, KernerKonhauser1993, AwRascle2000, Zhang2002, FlynnKasimovNaveRosalesSeibold2009, SeiboldFlynnKasimovRosales2013}. However, the focus of this study is on the equilibrium quantities only. When it comes to fuel consumption estimates (see \S\ref{subsec:Fuel_Consumption}), non-equilibrium effects that are present in highly congested flow are likely to amplify the impact of the fundamental ideas presented here (see the discussion in \S\ref{sec:Further_Benefits}).

At the position of bottlenecks, coupling conditions as in \cite{CocliteGaravelloPiccoli2005, LattanzioMauriziPiccoli2011} are provided. An important distinction of this work from existing models for MBs \cite{LattanzioMauriziPiccoli2011} is that in those, the evolution of the MB itself depends on the ambient traffic state, thus yielding coupled PDE--ODE models. In contrast, here the speed of the MB is a control parameter. Moreover, as we restrict to piecewise constant density profiles (as they naturally arise in LWR with bottlenecks), we are only employing MB speed profiles that are constant in time, while the control is active (see \S\ref{sec:Results}).

The idea to control traffic flow via MBs is important in the context of autonomous vehicles (AVs). Via vehicle-to-vehicle and/or vehicle-to-infrastructure communication, they will have fast access to non-local information, such as incidents and congestion downstream. Moreover, they can execute driving protocols very accurately, and thus can automatically adapt to microscopic small-scale deviations from the macroscopic large-scale traffic profile; better than any human driver could. And in contrast to traditional, costly, means of traffic control, such as ramp metering or variable speed limits, AVs will be in the traffic stream (in a few years) either way. Therefore, the cost of using them for traffic control may amount to as little as a compensation of their operator. Finally, because only a single MB is needed, the ideas developed here can in principle be applied in the near future when we will have very low AV penetration rates.

This paper is organized as follows. First we describe the modeling framework (\S\ref{sec:Modeling_Framework}), which describes the setup (\S\ref{subsec:Outline_Problem_Setup}), the dynamics of the bottlenecks (\S\ref{subsec:Dynamics_Bottlenecks}), the choice of a realistic FD (\S\ref{subsec:realistic_FD}), the used fuel consumption estimates (\S\ref{subsec:Fuel_Consumption}), and the formulas used to quantify savings in fuel (\S\ref{subsec:Quantifying_Savings}). Next, we present quantitative results (\S\ref{sec:Results}), including optimal fuel consumption reduction values (\S\ref{subsec:Effect_of_d}, \S\ref{subsec:Optimal_MB_Speed}), and the sensitivity of the outcomes with respect to important model parameters (\S\ref{subsec:Effect_of_beta}). The paper closes with a discussion of further benefits of the presented control strategies (\S\ref{sec:Further_Benefits}), and general conclusions (\S\ref{sec:Conclusions}).

\begin{figure}
\centering
  \subfloat[Initial traffic state ($t<t_0$)]{%
    \includegraphics[width=\textwidth]{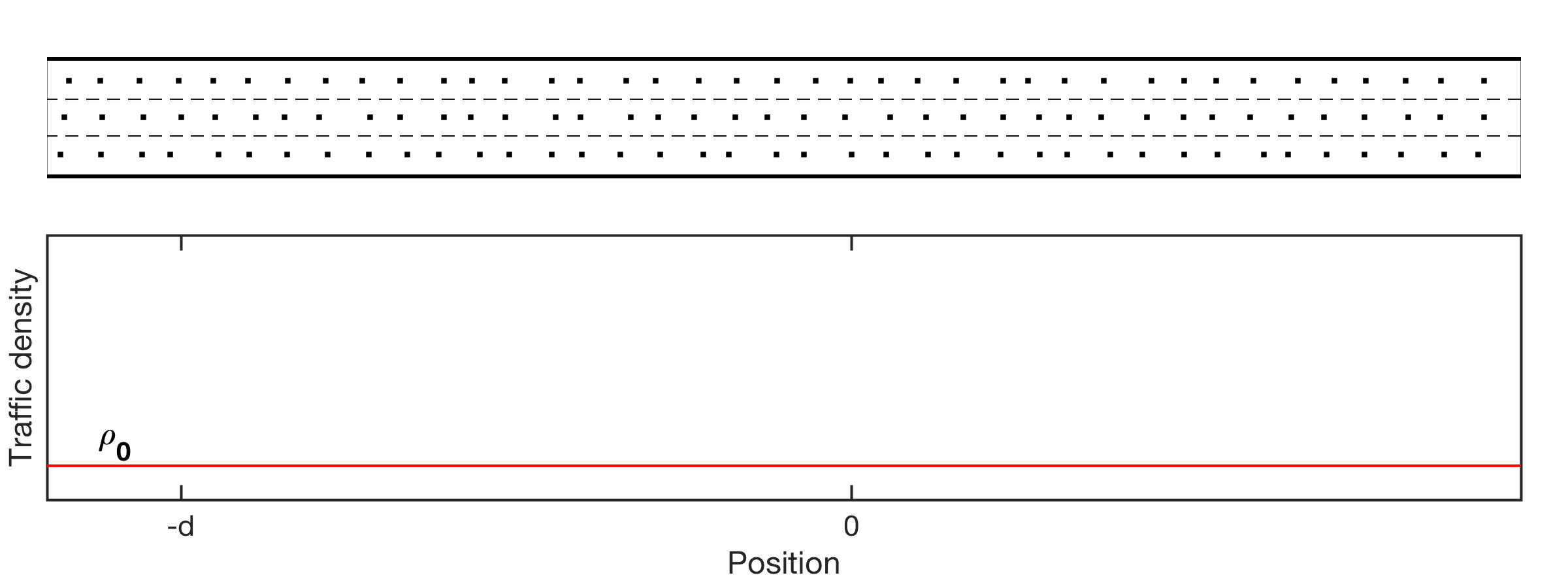}
}\\
  \subfloat[After the two bottlenecks have been activated ($t>t_1$)]{%
    \includegraphics[width=\textwidth]{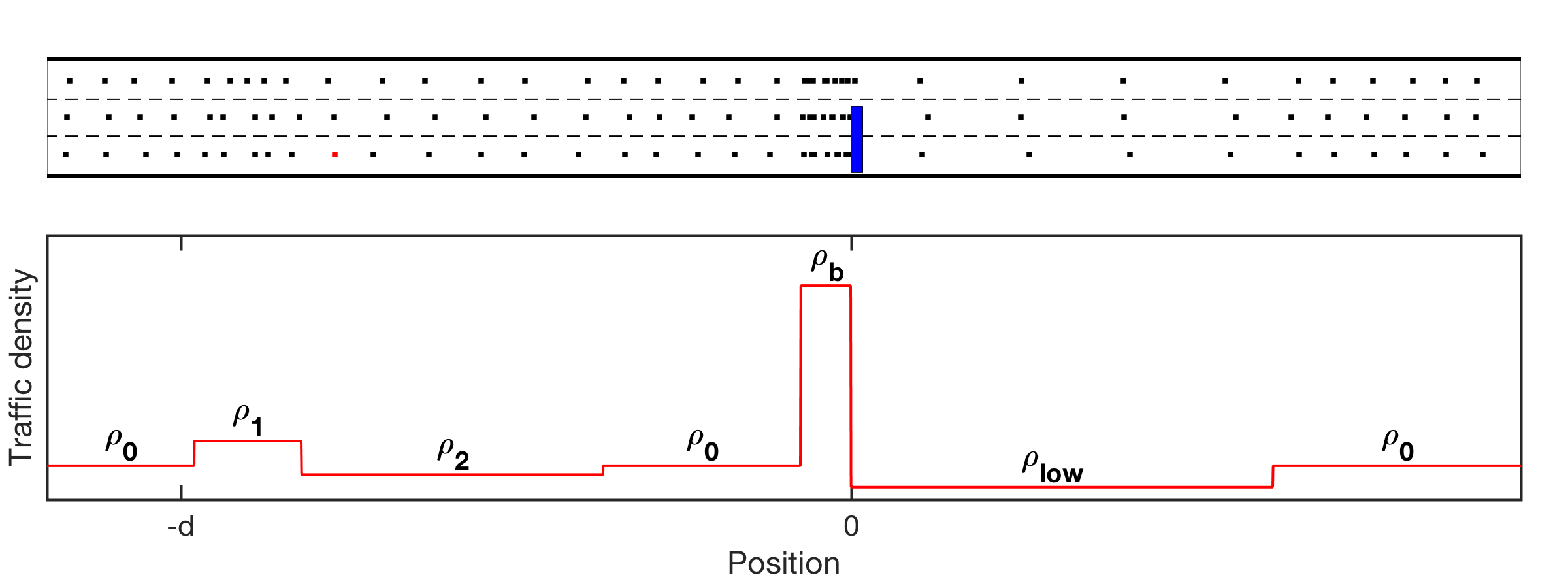}
}
\caption{Illustration of the problem setup: Panel~(a) shows the traffic state before the occurrence of the FB. Panel~(b) shows the traffic state after the FB (blue box) and the MB (red box) have been activated ($t>t_1$). In both panels, the density profile $\rho(\cdot,t)$ profile is augmented by a visualization in terms of actual vehicles (above).}
\label{fig:traffic_states}
\end{figure}

%===========================================================================
\vspace{1.5em}
\section{Modeling Framework}
\label{sec:Modeling_Framework}
%===========================================================================

%---------------------------------------------------------------------------
\subsection{Outline of the Problem Setup}
\label{subsec:Outline_Problem_Setup}
%---------------------------------------------------------------------------
The mathematical constructions described below hold for very general setups. However, to obtain meaningful estimates of the potential fuel consumption reduction generated by MBs, we make certain modeling choices that are close to real situations, however with some simplifications/idealizations. Specifically, we consider a highway (of infinite length) with 3 lanes, uniform road conditions, and no ramps. The jamming density is $\rho_\text{m}$. The initial traffic density is constant $\rho_0$. At time $t=t_0$, a FB (e.g., an accident) arises somewhere on the highway, closing down 2 lanes. The FB creates two new states on the highway: a state with high density $\rho_\text{b}>\rho_0$, traveling backwards in the form of a shock wave with constant velocity $s_\text{b}$, and a state with low density $\rho_\text{low}<\rho_0$, traveling forward in the form of a shock wave with constant velocity $s_\text{low}$.

At some time $t=t_1\ge t_0$, let $x=0$ be the position of the backwards moving shock produced by the FB. At that time $t=t_1$, we activate a MB at position $x=-d$, i.e., at a distance $d$ upstream from the high-density congestion, a vehicle in the right lane starts driving at a constant speed $s$, which is slower than the ambient equilibrium traffic flow. This MB creates two new states on the highway: a state with low density $\rho_2<\rho_0$ ahead of it, traveling forward at a constant velocity $s_2>s$, and a state with high density $\rho_1>\rho_0$ behind it, traveling at constant velocity $s_1<s$. This last state could be moving backwards or forward, depending on $\rho_0$ and the choice of $s$ (see \S\ref{subsec:Dynamics_Bottlenecks}). The traffic states before and after the activation of the two bottlenecks are depicted in Fig.~\ref{fig:traffic_states}, in terms of density profiles $\rho$ vs.~$x$, as well as visualized how the states can look in terms of vehicles.

\begin{figure}
\centering
  \subfloat[$t=t_1$: Time of MB activation at $x=-20$km]{%
   \setlength{\fboxsep}{0pt}%
    \setlength{\fboxrule}{1pt}%
    \fbox{\includegraphics[width=.45\textwidth]{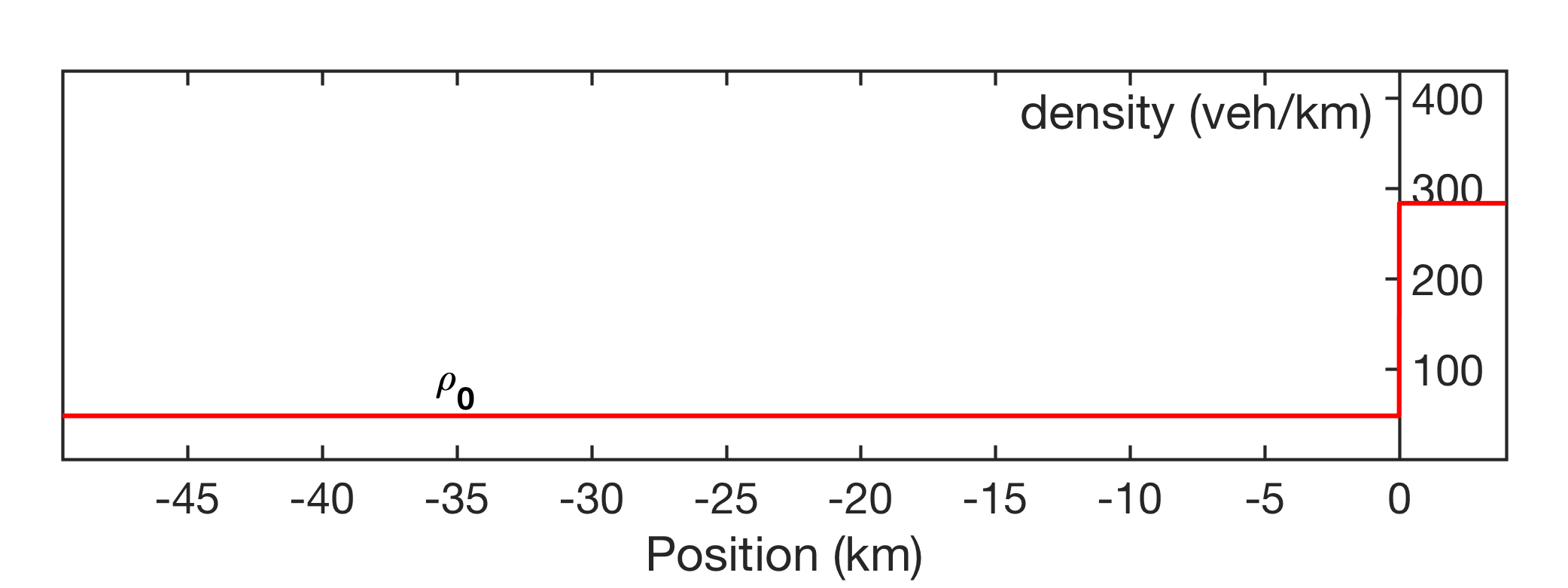}}
}\hfil
  \subfloat[$t=\frac{t_1+t_2}{2}$: Shortly after MB activation]{%
   \setlength{\fboxsep}{0pt}%
    \setlength{\fboxrule}{1pt}%
    \fbox{\includegraphics[width=.45\textwidth]{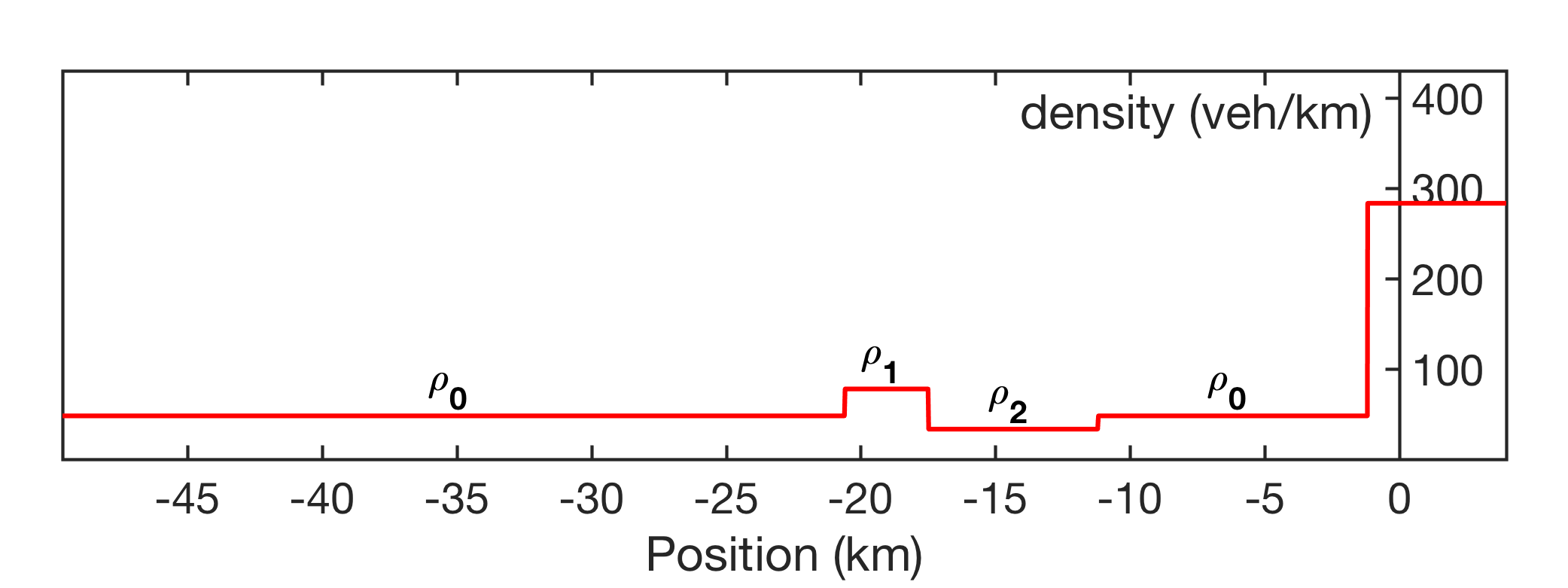}}
}\hfil
\\
  \subfloat[$t=t_2$: Time of first wave interaction]{%
   \setlength{\fboxsep}{0pt}%
    \setlength{\fboxrule}{1pt}%
    \fbox{\includegraphics[width=.45\textwidth]{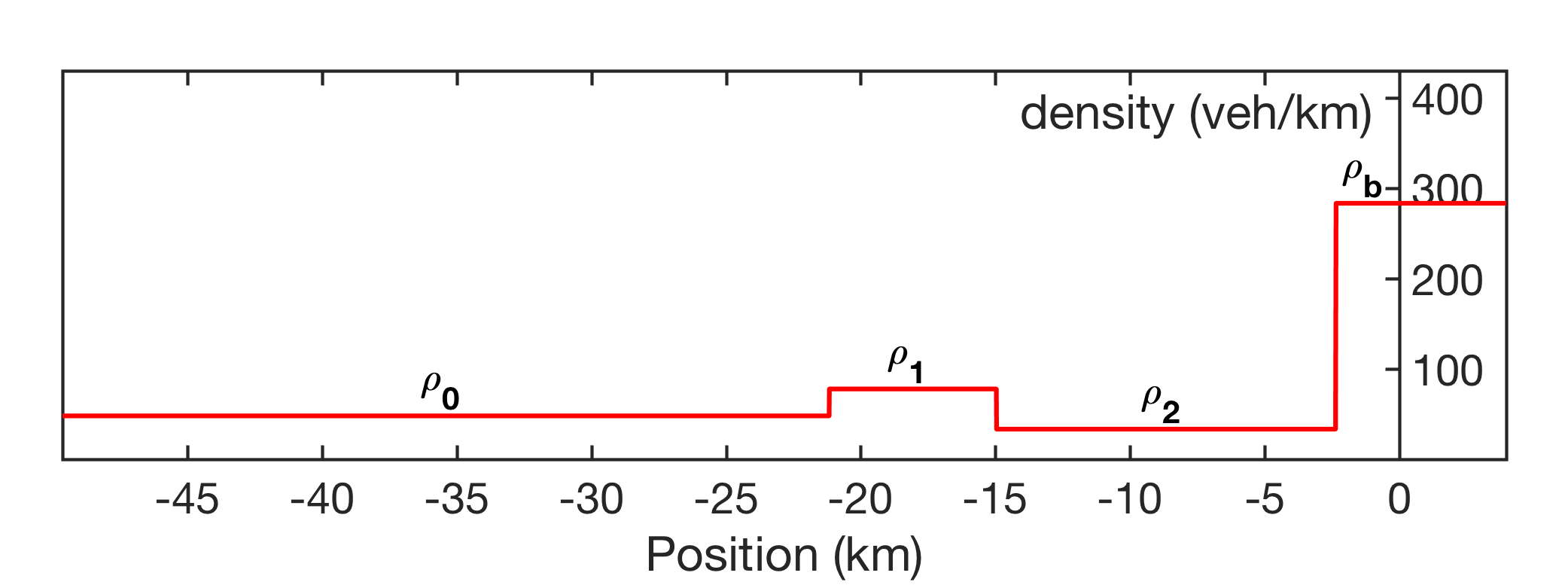}}
}\hfil
  \subfloat[$t=\frac{t_2+t_3}{2}$: Shortly after first wave interaction]{%
   \setlength{\fboxsep}{0pt}%
    \setlength{\fboxrule}{1pt}%
    \fbox{\includegraphics[width=.45\textwidth]{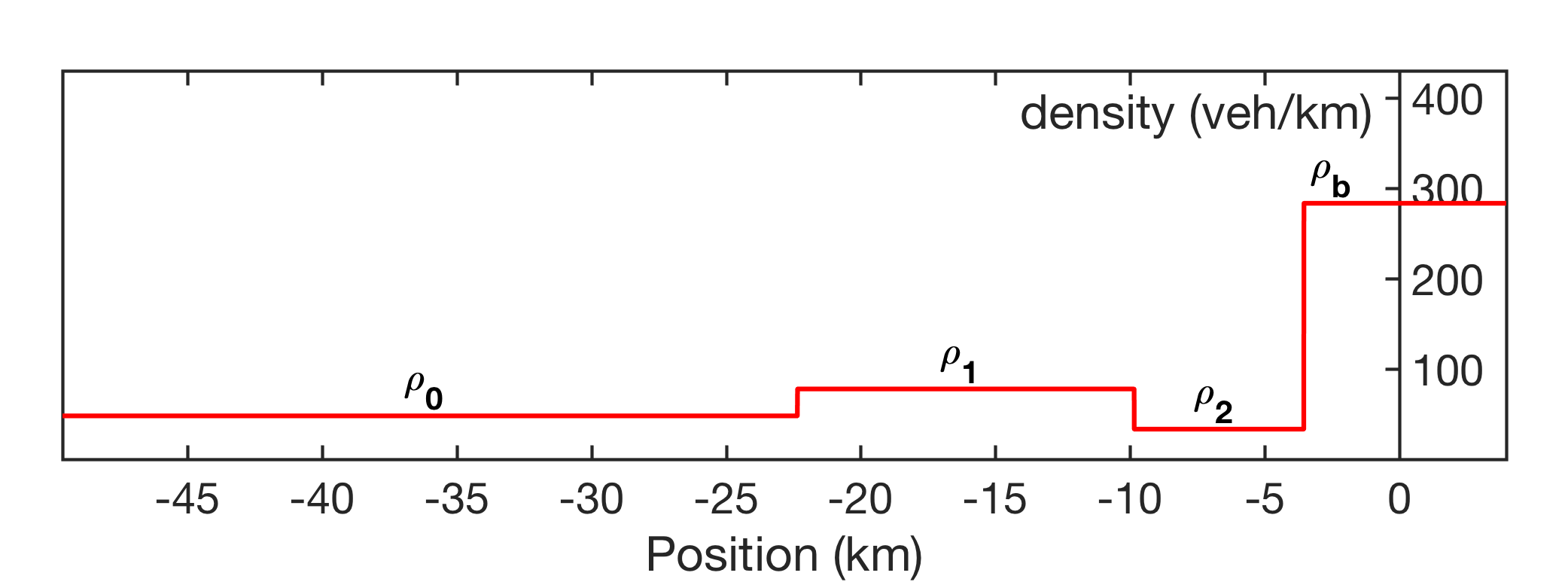}}
}\hfil
\\
  \subfloat[$t=t_3$: Time of second interaction \& MB deactivation]{%
   \setlength{\fboxsep}{0pt}%
    \setlength{\fboxrule}{1pt}%
    \fbox{\includegraphics[width=.45\textwidth]{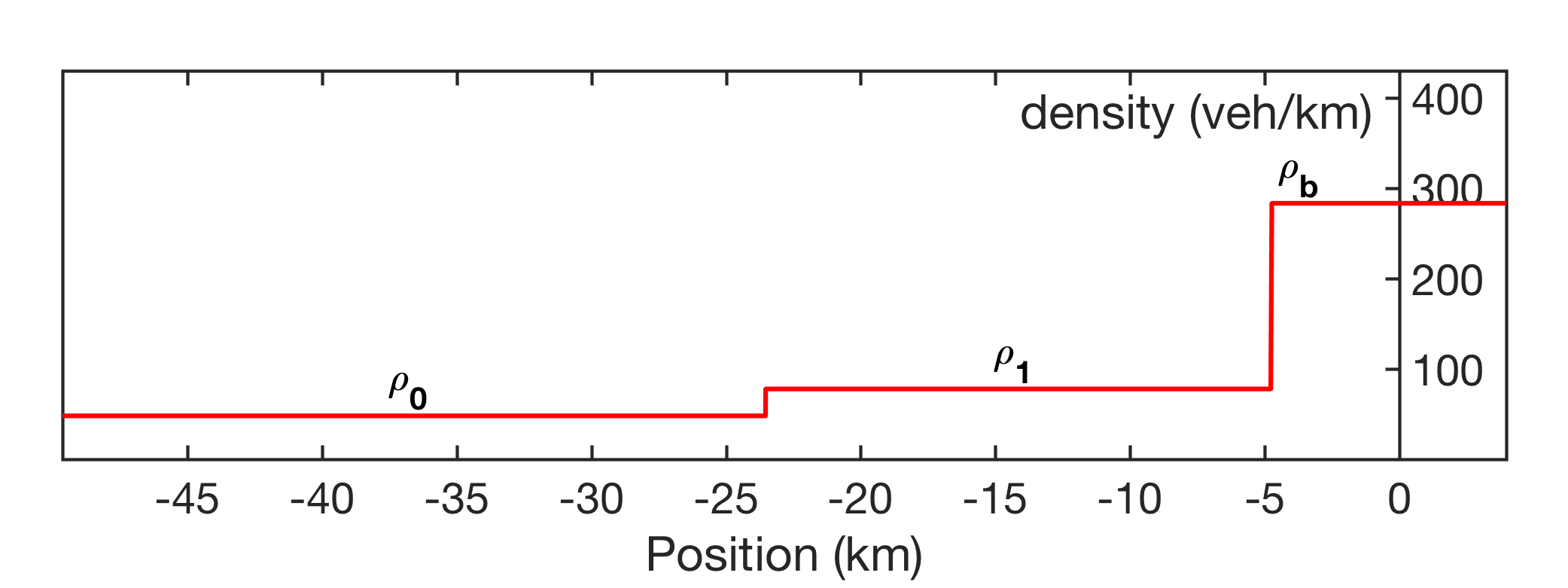}}
}\hfil
  \subfloat[$t=\frac{t_3+t_4}{2}$: Shortly after second wave interaction]{%
   \setlength{\fboxsep}{0pt}%
    \setlength{\fboxrule}{1pt}%
    \fbox{\includegraphics[width=.45\textwidth]{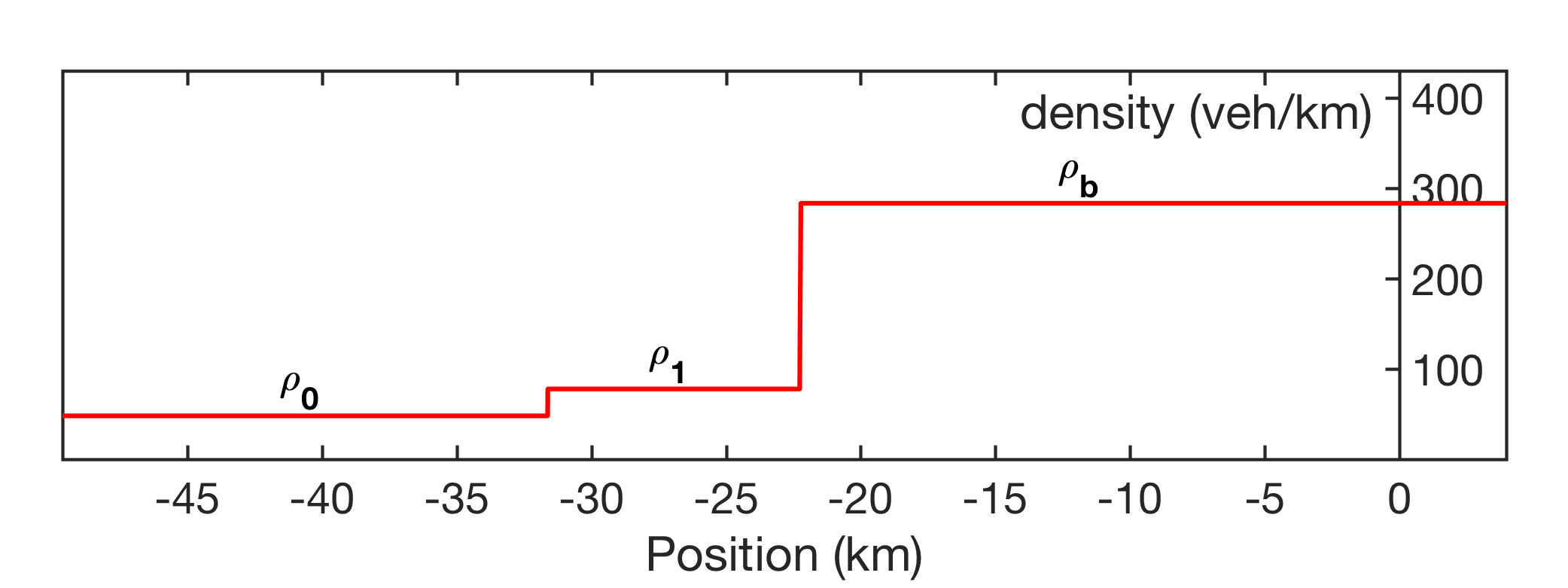}}
}\hfil
\\
  \subfloat[$t=t_4$: Time of third wave interaction]{%
   \setlength{\fboxsep}{0pt}%
    \setlength{\fboxrule}{1pt}%
    \fbox{\includegraphics[width=.45\textwidth]{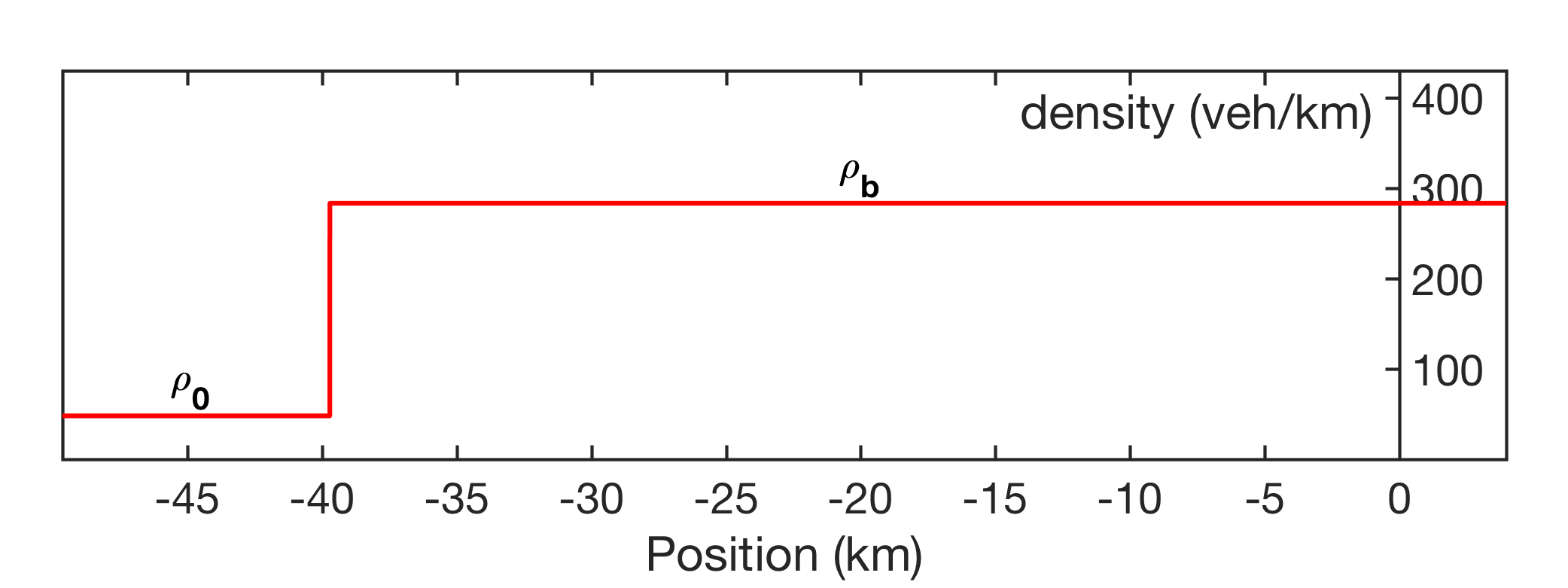}}
}\hfil
  \subfloat[$t=t_5 > t_4$: Effect of MB has vanished]{%
   \setlength{\fboxsep}{0pt}%
    \setlength{\fboxrule}{1pt}%
    \fbox{\includegraphics[width=.45\textwidth]{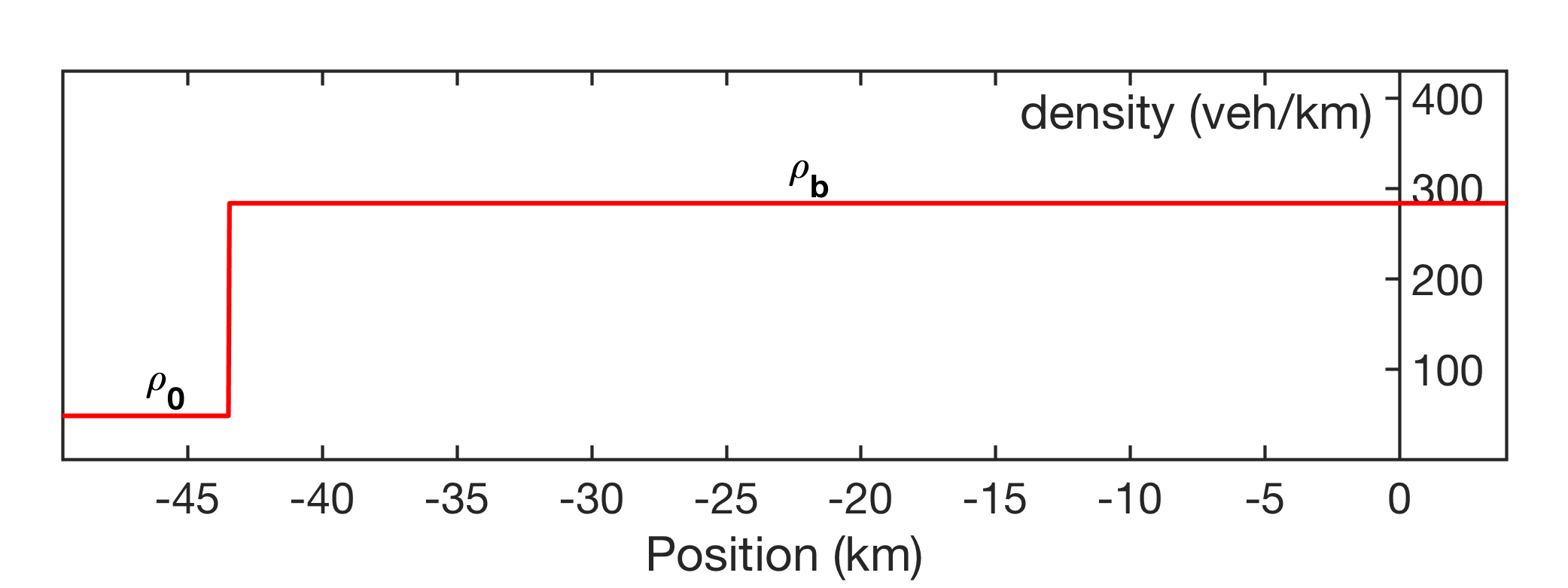}}
}\hfil
\caption{Snapshots of the temporal evolution of the traffic density profile on the highway at different times, from the activation of the MB until the effects of the MB have vanished.}
\label{fig:traffic_evolution}
\end{figure}

If we consider a concave downwards FD (such as the Greenshields flux), or if we consider a triangular FD and have $\rho_0$ in the free flow regime, the traffic waves produced by the FB and the MB interact three times. The temporal evolution of the traffic state on the highway, shown in Fig.~\ref{fig:traffic_evolution}, is as follows. The first shock wave interaction ($t=t_2$) happens when the low density state ahead of the MB meets the high density state upstream of the FB. This interaction results in a new shock wave that could either move forward or backwards, with constant velocity $s_3$. The second interaction ($t=t_3$) happens when the controlled vehicle (the MB) meets the traffic wave that has arisen from the first interaction. At this time $t=t_3$, we deactivate the MB (because it now enters the highly congested state $\rho_\text{b}$, in which there is no more point to slow down traffic). This second interaction results in a new shock travelling backwards at a constant velocity $s_4$. The third and final interaction ($t=t_4$) happens when the shock produced by the second interaction catches up to the high density state produced by the MB. After $t=t_4$, the influence of the MB has vanished, i.e., the traffic state is the same that it would have been without the activation of the MB. This implies that none of the vehicles on the highway have experienced any delay, except for the MB itself (which is negligible in a large-scale, macroscopic, perspective).

Below, we demonstrate that the introduction of the MB can have a positive impact (i.e., reduction) on the collective consumption of fuel by all the vehicles on the highway. Moreover, we determine the maximum possible reduction of fuel consumption with respect to the two control parameters allowed here: the MB speed $s$, and its initial distance $d$ from the the position of the backwards moving shock produced by the FB.

%---------------------------------------------------------------------------
\subsection{Bottleneck Dynamics}
\label{subsec:Dynamics_Bottlenecks}
%---------------------------------------------------------------------------
The theory presented in this section applies to generic concave down flux functions, including piecewise linear fluxes. We use the Greenshields flux $Q(\rho) = \rho\,u_\text{m}(1-\rho/\rho_\text{m})$ to visualize the theory and geometric constructions, as it is representative of a general concave down flux function. Here $\rho_\text{m}$ is the jamming density, and $u_\text{m}$ is the maximum speed. Quantitative estimates will then be obtained using a data-fitted piecewise linear flux.

\begin{figure}
\centering
  \subfloat[Equilibrium states around a FB]{%
    \includegraphics[width=.49\textwidth]{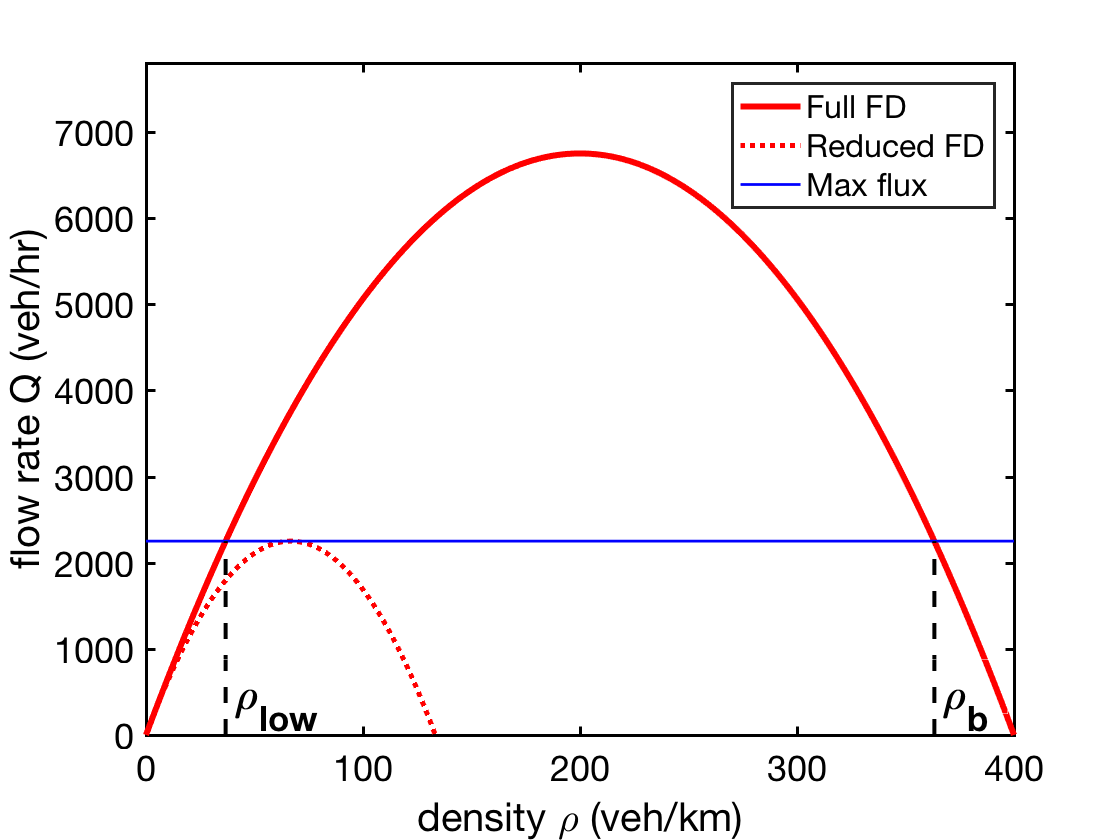}
}\hfill
  \subfloat[Equilibrium states around a MB]{%
    \includegraphics[width=.49\textwidth]{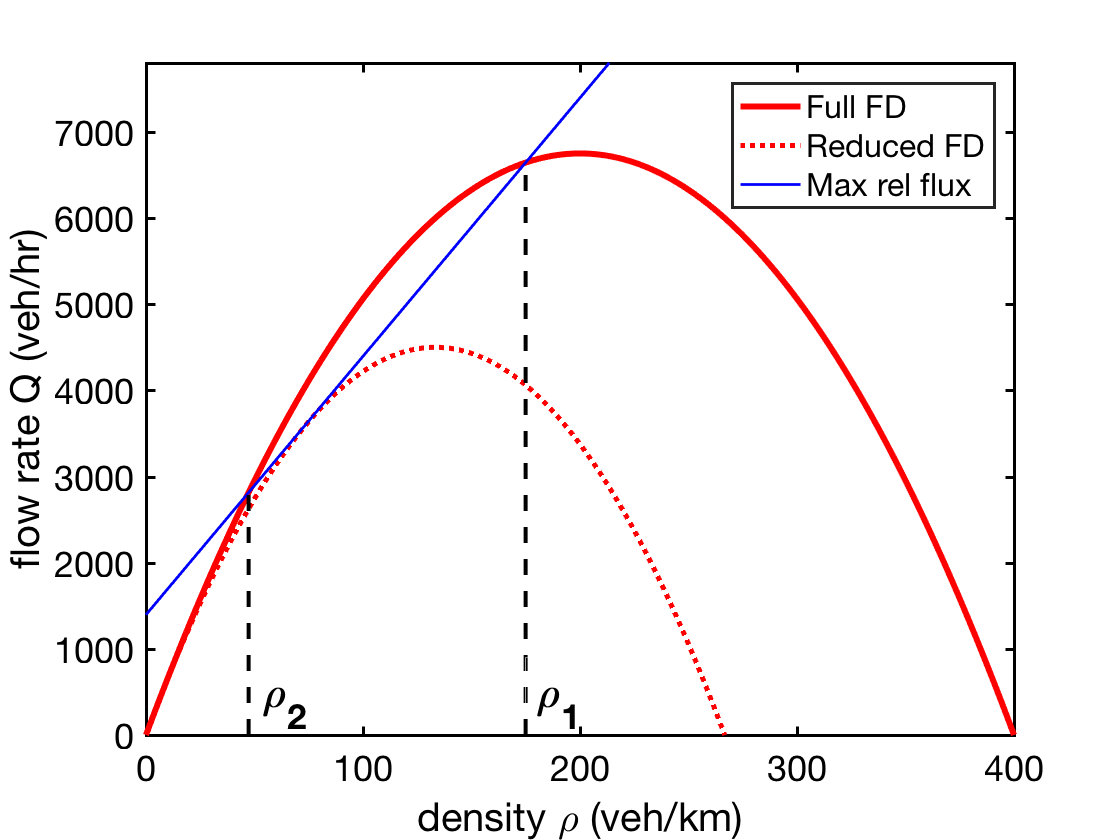}
}
\caption{Illustration (using the Greenshields flux) of the methodology to calculate the equilibrium density states around (a) a fixed bottleneck and (b) a moving bottleneck. In both cases, the maximum (relative) flux (blue line) with respect to the reduced FD (dashed red curve) is determined, and the intersection of that line with the full road FD (thick red curve) yields the newly generated traffic states.}
\label{fig:bottleneck_construction}
\end{figure}

\subsubsection{Fixed bottleneck}
We consider a highway with 3 lanes, with the FB blocking 2 lanes. The maximum flux allowed past the FB is one third\footnote{Reduced passing speeds would yield even lower maximum flux; the construction works similarly.} of the original maximum flux $Q_\text{m} = \frac{u_\text{m}\rho_\text{m}}{4}$. If $Q(\rho_0)<\frac{Q_\text{m}}{3}$, the FB has no effect on the traffic density of the highway (all vehicles can pass). In turn, if $Q(\rho_0)>\frac{Q_\text{m}}{3}$, two new density states are produced, corresponding to the two roots of the relationship $Q(\rho) = \frac{Q_\text{m}}{3}$, given by the intersection of the blue line with the thick red curve in Fig.~\ref{fig:bottleneck_construction}, panel~(a). Let $\rho_\text{b}$ denote the high density state upstream of the FB, and $\rho_\text{low}$ the low density state ahead of the FB. The shock at the downstream end of the low density region moves forward with velocity $s_\text{low} = (Q(\rho_\text{low})-Q(\rho_0))/(\rho_\text{low}-\rho_0)$. Likewise, the shock at the upstream end of the high density region moves backwards, with velocity $s_\text{b} = (Q(\rho_\text{b})-Q(\rho_0))/(\rho_\text{b}-\rho_0)$.

\subsubsection{Moving bottleneck}
The effect of a MB on the macroscopic traffic state is similar to a FB, however with fluxes considered in a moving frame of reference. In particular, a MB does not always affect the density profile along the highway. If the density $\rho_0$ is sufficiently low, all vehicles can pass the MB and the traffic state remains unchanged. In turn, if traffic is sufficiently dense (the regime in which traffic control is largely of interest), the presence of a MB produces a higher density state in its wake, while producing a lower density state ahead of itself.

We consider the MB to occupy one out of the three lanes of the highway. The fundamental diagram corresponding to the remaining two lanes at the position of the MB is $\beta Q(\rho/\beta)$, where $0<\beta<1$ is the reduction factor of the original fundamental diagram of the highway with three lanes. The most natural choice for the situation at hand is $\beta = \frac{2}{3}$; however, lower values could arise due to lane-changing ``friction'' effects caused by the MB (see \S\ref{subsec:Effect_of_beta}).

Given a flow rate curve $Q(\rho)$, the flux relative to a frame of reference moving with speed $s$ is $Q(\rho)-s\rho$. Therefore, the maximum flux past the MB is $Q^{\text{rel}} = \beta Q(\rho/\beta)-s\rho$. Geometrically (see Fig.~\ref{fig:bottleneck_construction}, panel~(b)), the maximum relative flux $Q^{\text{rel}}_{\text{m}}$ is the intercept of the line with slope $s$ (blue) that is tangent to the reduced FD $\beta Q(\rho/\beta)$ (dashed red).

As the MB is driving at a constant speed $s$, the high ($\rho_1$) and low ($\rho_2$) density states produced by the MB correspond to two points on the FD $Q(\rho)$ connected by a line with slope $s$. To maximize $Q^{\text{rel}}$, these two points are the intersections of $Q(\rho)$ and the maximal relative flux line (blue line in Fig.~\ref{fig:bottleneck_construction}, panel~(b)). Note that with a MB, it is possible that the high density state $\rho_1$ is actually in the free flow regime. This is impossible with a FB. Note further that the densities $\rho_1$ and $\rho_2$ are independent of $\rho_0$. That base density $\rho_0$ affects only \emph{whether} the MB has an effect or not (if $\rho_0\ge\rho_1$ or $\rho_0\le\rho_2$, the MB has no effect).

In turn, if $\rho_1<\rho_0<\rho_2$, the MB produces a high density state ($\rho_1$) behind it, whose upstream shock travels with velocity $s_1 = (Q(\rho_1)-Q(\rho_0))/(\rho_1-\rho_0)$, and a low density state ($\rho_2$) ahead, whose downstream shock travels with velocity $s_2 = (Q(\rho_2)-Q(\rho_0))/(\rho_2-\rho_0)$. The concavity of $Q(\rho)$ implies that $s_1<s<s_2$.

\begin{figure}
\centering
\includegraphics[scale=0.25]{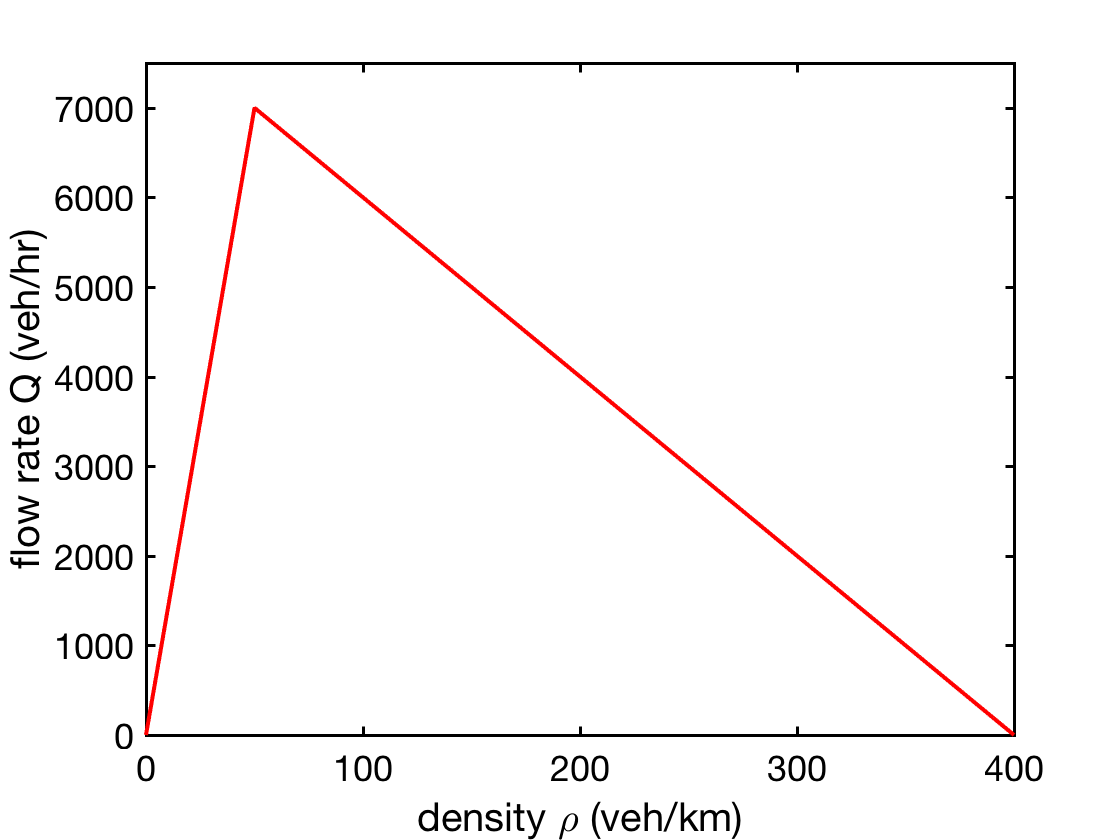}
\caption{The Newell-Daganzo FD used for the quantitative results.}
\label{fig:FD_data_fitted}
\end{figure}

%---------------------------------------------------------------------------
\subsection{Data-Fitted Fundamental Diagram}
\label{subsec:realistic_FD}
%---------------------------------------------------------------------------
In order to obtain quantitative estimates that reflect the overall fuel consumption of real traffic flow, we consider a data-fitted triangular piecewise linear Newell--Daganzo \cite{Newell1993, Daganzo1994} flux function throughout the rest of this paper:
\begin{equation}
\label{eq:NDFD}
\ Q(\rho) =
  \begin{cases}
       Q_{\text{m}} \frac{\rho}{\rho_{\text{c}}}    \hfill & \text{ for $ 0\leq \rho \leq \rho_{\text{c}} $ } \\
       Q_{\text{m}} \frac{\rho_{\text{m}}-\rho}{\rho_\text{m}-\rho_{\text{c}}}    \hfill & \text{ for $ \rho_\text{c}\leq \rho \leq \rho_{\text{m}} $}\;. \\
  \end{cases}
\
\end{equation}
Here $\rho_\text{c}$ is the critical density at which the maximum flux, $Q_{\text{m}}$, is achieved. We choose
\begin{equation*}
\ \rho_{\text{m}}=\frac{\#\text{lanes}}{7.5\text{m}} = \frac{3}{7.5\text{m}} = 400\> \text{veh/km}\;,
\end{equation*}
where the 7.5m represent 5m average vehicle length, plus 50\% safety distance, as justified in \cite{FanSeibold2013}. Moreover, motivated by German highways \cite{wu2002new}, we choose the free flow speed $u_\text{m} = 140$ km/hr, and $\rho_\text{c}=\frac{\rho_\text{m}}{8} = 50$ veh/km. The resulting maximum flux is $Q_\text{m}=u_\text{m} \rho_\text{c}= 7000$ veh/hr. Figure~\ref{fig:FD_data_fitted} shows the graph of the FD with these values.

%---------------------------------------------------------------------------
\subsection{Fuel Consumption}
\label{subsec:Fuel_Consumption}
%---------------------------------------------------------------------------
Our aim is to quantify the overall fuel consumption (FC) of all vehicles on the highway in various situations. The relationship between the speed of a vehicle and its FC efficiency (Liters/km) is discussed in \cite{ahn2002estimating, Berry2010}: vehicles consume more fuel per distance traveled when they are driving at very low speeds (a certain amount of fuel is used to just keep the engine, and accessories, running) or at very high speeds (more energy is needed to overcome air drag).

\begin{figure}
\centering
  \subfloat[Vehicle FC rate $K(s)$ vs.~speed]{%
    \includegraphics[width=.49\textwidth]{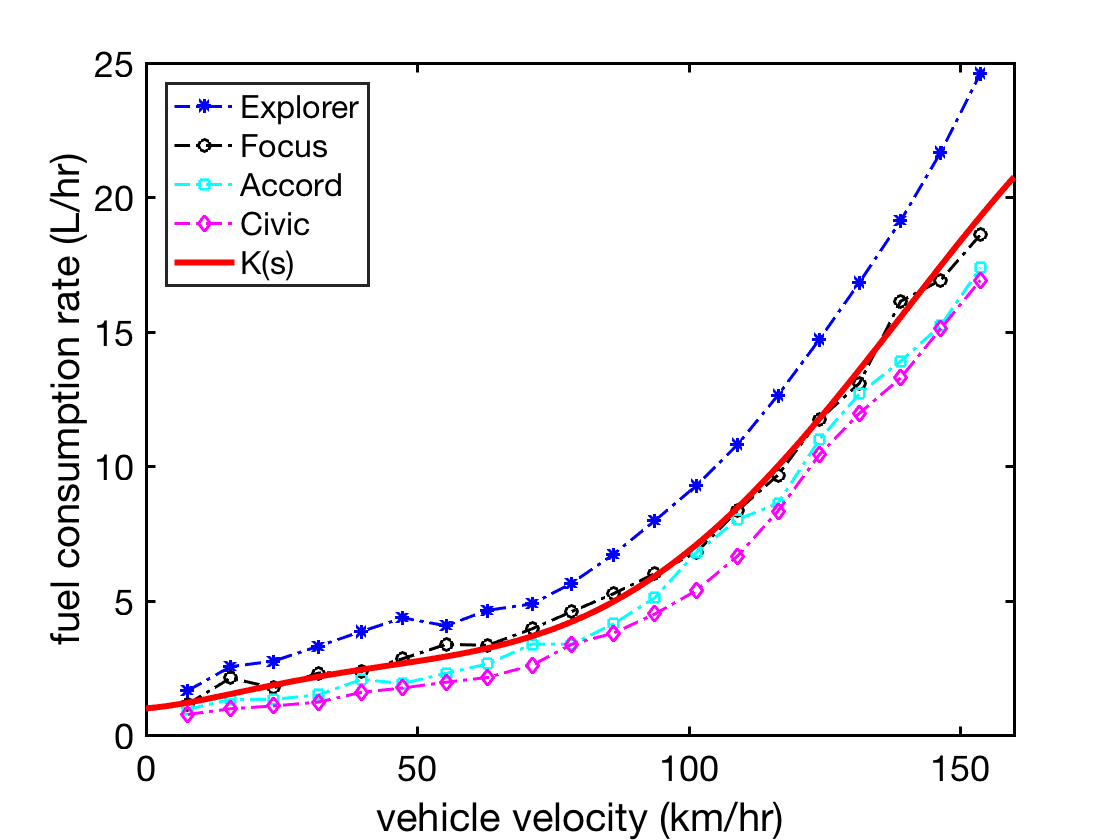}
}\hfill
  \subfloat[Total FC rate $F(\rho) = \rho K(U(\rho))$ vs.~density]{%
    \includegraphics[width=.49\textwidth]{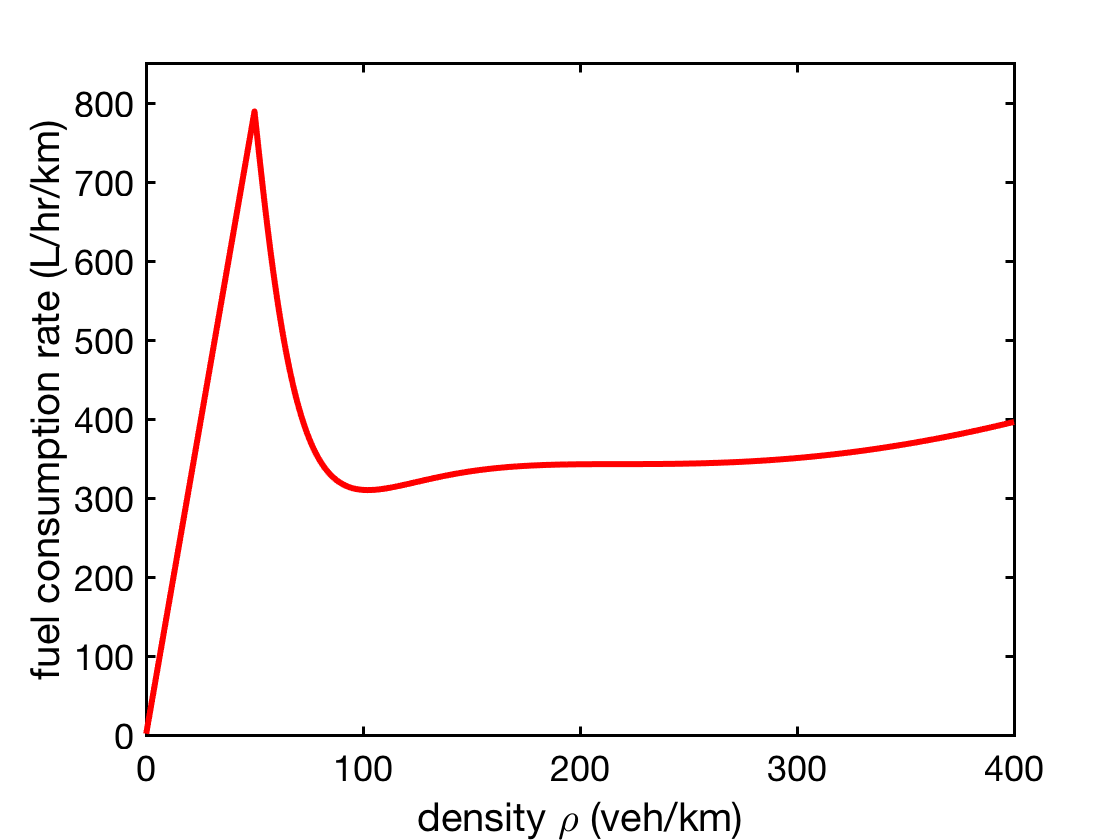}
}
\caption{Plots of the fuel consumption rate functions. Panel~(a) shows the FC rate of the four vehicles considered in \cite{Berry2010} vs.~the vehicle velocity, and the fitted average polynomial $K(s)$ (solid red curve). Panel~(b) shows the density-dependence of FC rate $F(\rho) = \rho K(U(\rho))$ of all vehicles per unit length.}
\label{fig:fuel_consumption}
\end{figure}

Figure~41 in \cite{Berry2010} provides fuel consumption efficiency data (Liters/km) for four types of vehicles (Ford Explorer, Ford Focus, Honda Civic, and Honda Accord), as functions of the vehicle speed. Multiplying the FC efficiency by the vehicle speed yields the FC rate (Liters/hr). The FC rates for the four vehicles as functions of the vehicle speed are shown in Fig.~\ref{fig:fuel_consumption}, panel~(a). We average these four curves (with equal weights), and approximate (in a least squares sense) the resulting data points via a sixth order polynomial (which is accurate up to a 5\% error). This average FC rate function reads as:
\begin{equation}
\label{eq:FC_rate_vs_speed}
K(s) = 5.7\times10^{-12}s^6 - 3.6\times10^{-9}s^5 + 7.6\times10^{-7}s^4 - 6.1\times10^{-5}s^3 + 1.9\times10^{-3}s^2 + 1.6\times10^{-2}s + 0.99\;.
\end{equation}
Here units have been omitted for notational efficiency ($s$ is in km/hr and $K$ is in Liters/hr). The function is given by the red curve in Fig.~\ref{fig:fuel_consumption}, panel~(a).

Assuming the LWR model \eqref{eq:LWR}, we can now quantify the FC rate of the whole traffic, as a function of the traffic density. To that end, we (a) re-parametrize \eqref{eq:FC_rate_vs_speed} in terms of the density $\rho$, as given by the LWR model; and (b) multiply $K$ by $\rho$ to obtain the FC rate \emph{of traffic}, rather than of \emph{a single vehicle}. Considering the triangular FD \eqref{eq:NDFD} from \S\ref{subsec:realistic_FD}, the bulk velocity vs.~density relationship is
\begin{equation*}
\begin{split}
U(\rho) &=
\begin{cases}
       u_\text{m}    \hfill & \text{ for $ 0\leq \rho \leq \rho_{\text{c}} $ } \\
       u_{\text{m}} (\frac{\rho_\text{c}}{\rho}) (\frac{\rho_{\text{m}}-\rho}{\rho_\text{m}-\rho_{\text{c}}})    \hfill & \text{ for $ \rho_\text{c}\leq \rho \leq \rho_{\text{m}} $ } \\
  \end{cases} \\[.2em]
&= \begin{cases}
       140\>\text{km/hr}    \hfill & \text{ for } \rho \leq \text{50 veh/km} \\
       \frac{8000/\text{hr}}{\rho} - 20\>\text{km/hr}  \hfill
       & \text{ for } \rho \geq \text{50 veh/km}\;. \\
  \end{cases}
\end{split}
\end{equation*}
The FC rate (Liters/hr) of one vehicle vs.~the traffic density at the vehicle's position is then $f(\rho) = K(U(\rho))$. And the total FC rate (Liters/hr) of a segment of the highway vs.~the total traffic density of this segment is given by $F(\rho) = \rho f(\rho)$. Figure~\ref{fig:fuel_consumption}, panel~(b) shows this function. It is linear is $\rho$ when $0 \leq \rho \leq \rho_\text{c}$, because in the free flow regime the speed $U(\rho) = u_\text{m}$ is constant. Moreover, it reaches its maximum at the critical density.

%---------------------------------------------------------------------------
\subsection{Quantifying Savings in Fuel}
\label{subsec:Quantifying_Savings}
%---------------------------------------------------------------------------
Based on the FC rate function $F(\rho)$, constructed above, we can quantify the impact of a MB on the total FC of all vehicles on a segment of the highway. The following two scenarios are being compared:
\begin{itemize}
\item \textit{Scenario A:} The FB arises, but the MB is not activated (uncontrolled case).
\item \textit{Scenario B:} The FB arises, and the MB is activated (controlled case).
\end{itemize}
Because the effect of the MB vanishes after some finite time (see \S\ref{sec:Modeling_Framework}), both scenarios return to the same traffic state eventually, however via different traffic state evolutions before that.

\begin{figure}
\centering
  \subfloat[The ``local'' domain $\Omega$]{%
    \includegraphics[width=.45\textwidth]{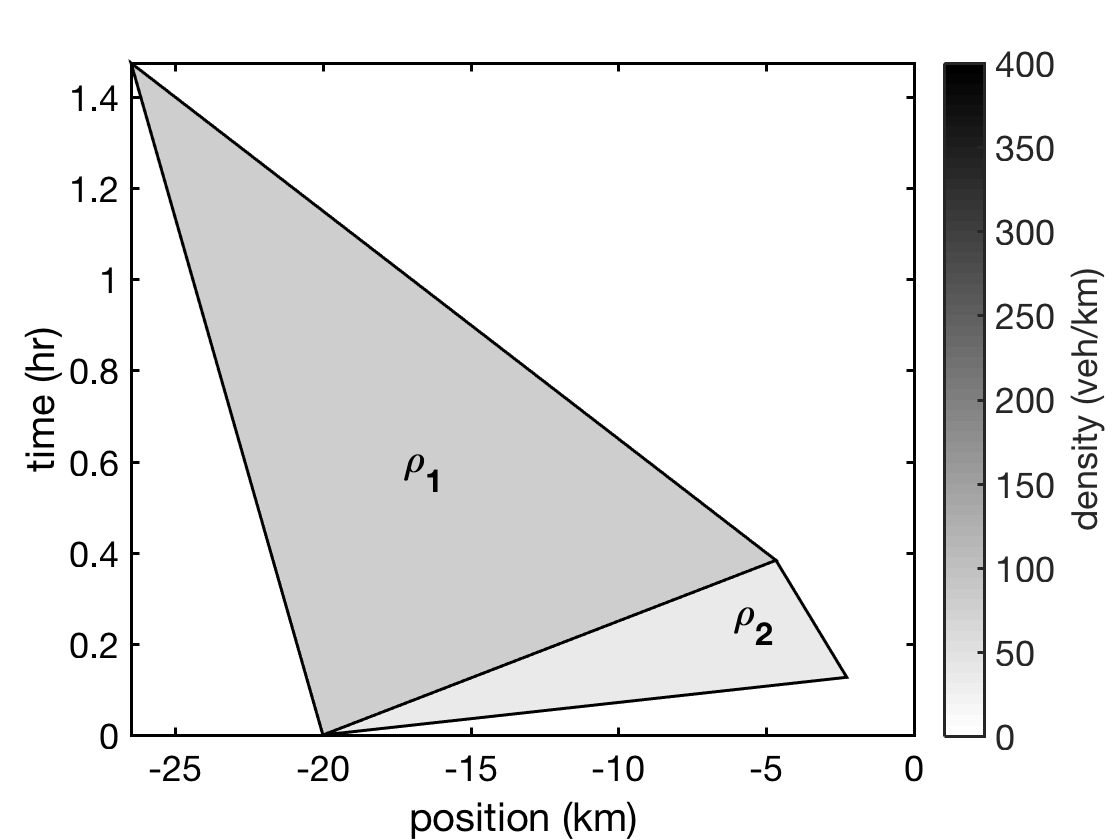}
}\hspace{0.2in}
  \subfloat[The ``global'' domain $(z,0)\times(t_1,t_4)$]{%
    \includegraphics[width=.45\textwidth]{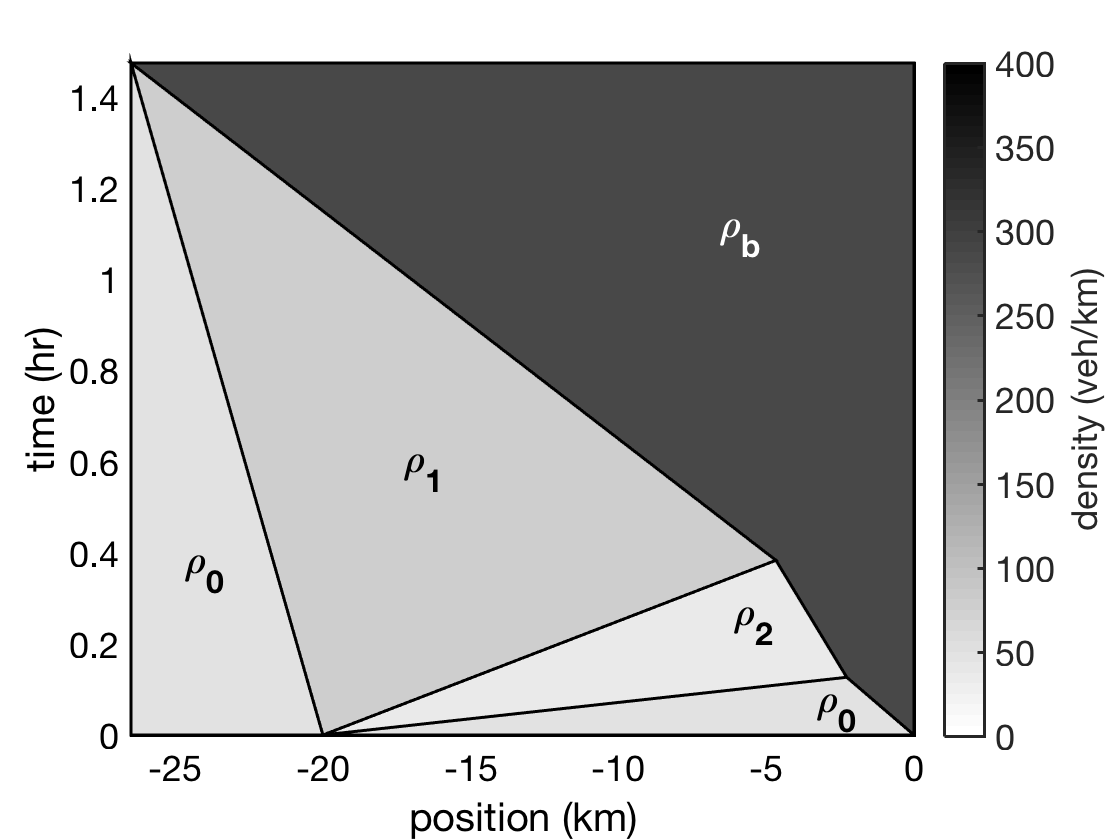}
}\hfill
\caption{Domains (in the $(x,t)$-plane) used for quantifying the FC reduction. Panel~(a) shows the ``local'' domain $\Omega$, which is the region of influence of the MB. Panel~(b) shows the ``global'' domain $(z,0)\times(t_1,t_4)$, which is the smallest rectangle in the $(x,t)$-plane that includes $\Omega$.}
\label{fig:domains}
\end{figure}

\subsubsection{Absolute savings in fuel}
We consider the total amount of fuel that is saved per hour due to the implementation of the controlled scenario~B, rather than simply letting scenario~A unfold. Let $\rho_A(x,t)$ be the density on the highway in scenario $A$, and $\rho_B(x,t)$ be the density on the highway in scenario~B. Due to the finite speed of information propagation in the LWR model \eqref{eq:LWR}, a MB can only affect the traffic state in its vicinity (in space-time). It cannot affect the traffic density on all of the highway, and its effect is limited to a finite interval in space and in time ($t\in [t_1,t_4]$). Specifically, we define
\begin{equation*}
%\label{eq:omega}
\Omega := \{(x,t)\mid \rho_\text{A}(x,t) \neq \rho_\text{B}(x,t)\}
\end{equation*}
to be the domain of influence of the MB (shown in Fig.~\ref{fig:domains}, panel~(a)). The total FC of traffic in $\Omega$ is
\begin{equation}
\label{eq:Gomega}
G_X^{\Omega} = \iint\limits_{\Omega} F(\rho_X(x,t)) \,\text{d}x\,\text{d}t\;,
\quad\quad X \in \{\text{A},\text{B}\}\;.
\end{equation}
Therefore, the total fuel saved due to the MB is
\begin{equation}
\label{eq:W}
W = G_\text{A}^{\Omega}-G_\text{B}^{\Omega}\;,
\end{equation}
and we measure the \emph{fuel saving rate}
\begin{equation*}
%\label{eq:U}
Y = \frac{W}{t_4-t_1}
\end{equation*}
as the total amount of fuel saved (due to the MB), divided by the duration of influence of the MB.

\subsubsection{Relative savings in fuel}
A relative measure of the impact of the MB control on the fuel consumed is obtained by dividing the total fuel saved \eqref{eq:W} by the total fuel consumed by the traffic flow. The challenge in this notion is to define which vehicles (and when) should be included in that ``total''. Clearly, it only makes sense to incorporate vehicles in the vicinity (in space-time) of the MB. Below we describe two possible notions of a ``total'' fuel consumption.

We define the \emph{local relative FC savings} as
\begin{equation}
\label{eq:Romega}
R^{\Omega} = 1- \frac{G_\text{B}^{\Omega}}{G_\text{A}^{\Omega}}\;,
\end{equation}
which is the total fuel saved (by the control) relative to the total fuel consumed by traffic on the domain of influence of the MB, $\Omega$, defined above. Note that (i) the size and shape of the domain $\Omega$ depends on the control parameters; and (ii) the segment of highway that is considered via $\Omega$ changes in time (see Fig.~\ref{fig:domains}, panel~(a)).

A geometrically simpler reference domain can be defined as follows. Consider the segment of the highway $x\in [z,0]$ where $z = \min(-d,s_\text{b} t_4)$, over the time interval $t\in [t_1,t_4]$. This rectangle in space-time (shown in Fig.~\ref{fig:domains}, panel~(b)) represents the portion of highway that is affected by the MB at any time, and the time interval from the MB's activation until its effect has vanished. By construction, this domain includes the domain of influence $\Omega$. Analogous to \eqref{eq:Gomega}, the total FC of traffic (in scenario~A or~B) over this rectangular domain is
\begin{equation*}
%\label{eq:G}
G_X = \int_{t_1}^{t_4} \int_{z}^{0}F(\rho_X(x,t)) \,\text{d}x\,\text{d}t\;,
\quad\quad X \in \{\text{A},\text{B}\}\;,
\end{equation*}
and the \emph{global relative FC savings} are given by
\begin{equation*}
%\label{eq:R}
R = 1- \frac{G_\text{B}}{G_\text{A}}\;.
\end{equation*}

Clearly, the quantity \eqref{eq:Romega} leads to the largest values of relative FC savings, as the domain $\Omega$ is the smallest reasonable reference region. In contrast, the larger rectangular domain $(z,0)\times(t_1,t_4)$ leads to smaller estimates of relative FC savings. A rationale for considering that larger domain (besides geometric simplicity) is that this region in space-time marks a fixed segment of highway that is under ``MB control''. Due to this structure, $R$ is a conservative measure of the relative impact of the MB, while $R^{\Omega}$ provides an upper bound on the effect of the MB in terms of relative fuel savings. The true relative impact of the MB is somewhere between $R$ and $R^{\Omega}$.

The quantities $Y$, $R$, and $R^{\Omega}$ depend on the parameters $\rho_0$, $s$, and $d$. Therefore, given an initial traffic density $\rho_0$, we can determine for which choices of the control parameters $s$ and $d$ the effect of the MB on FC reduction is maximized (with respect to $Y$, $R$, or $R^{\Omega}$). Note that, because of the macroscopic description, the analysis neglects the FC of the MB itself. That is justified, because on a large highway segment, its contribution is insignificant relative to the bulk flow.

\begin{figure}
\centering
  \subfloat[$d=20$ km]{%
    \includegraphics[width=.45\textwidth]{GlobalDomain}
}\hspace{0.2in}
  \subfloat[$d=10$ km]{%
    \includegraphics[width=.45\textwidth]{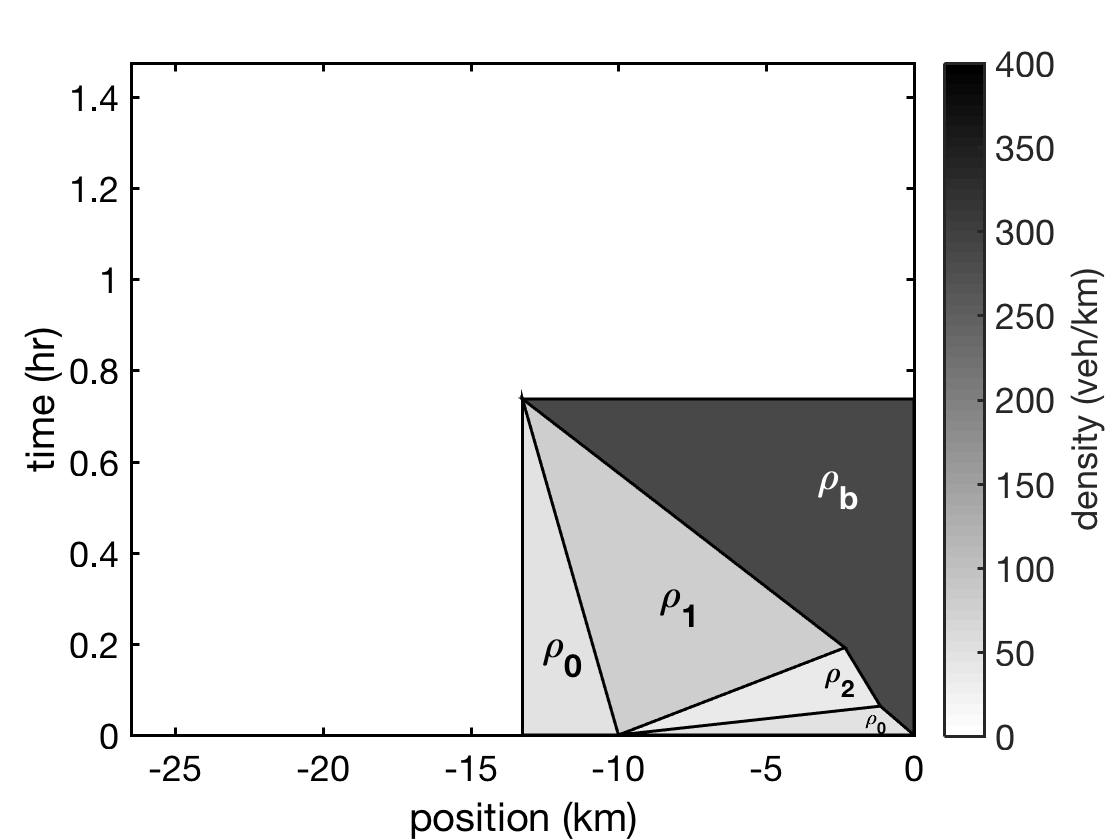}
}\hfill
\caption{Evolution of traffic density in the global domain of the $(x,t)$ space for $\rho_0=47$ veh/km, $s=40$ km/hr, and two different values of $d$. Panel~(a) has $d=20$ km, and panel~(b) has $d=10$ km. One can see that the solution in the $(x,t)$-domain becomes re-scaled with $d$ in both $x$ and $t$.}
\label{fig:domain_rescale}
\end{figure}

%===========================================================================
\vspace{1.5em}
\section{Results}
\label{sec:Results}
%===========================================================================

%---------------------------------------------------------------------------
\subsection{Effect of Distance $d$}
\label{subsec:Effect_of_d}
%---------------------------------------------------------------------------
Although only one MB is needed to implement the suggested control, multiple AVs may be on the highway that could act as potential control vehicles. For a 3-lane density of $\rho_0 =$ 50 veh/km, an AV penetration rate of merely 4\% implies that, on average, every 500m an AV can be found. Therefore, if AVs are roughly evenly distributed along the highway, a desired distance $d$ between the congestion zone (caused by the FB) and the activated MB can be realized up to 250m precision. Under this premise, the distance $d$ becomes a control parameter.

We aim to choose $d$ such that fuel savings are maximized. As a matter of fact, the distance $d$ exactly re-scales the evolution of the density on the highway with respect to both $x$ and $t$, as shown in Fig.~\ref{fig:domain_rescale}. This is evident by the fact that all speeds $s_1$, $s_2$, $s_\text{b}$, and $s_\text{low}$, as well as the function $F(\rho)$, are independent of $d$. Therefore, the absolute fuel savings rate $Y$ scales linearly with $d$, i.e., $Y(\lambda d) = \lambda Y(d)$ for $\lambda>0$. This implies that if we were to activate only a single MB on the highway, it is advisable to maximize $d$, as long as the effects of that MB will have vanished by the time the FB clears. This last requirement is important: any vehicles held back by the MB that do not hit the high density state $\rho_\text{b}$ anymore (because the FB has been cleared), will experience an actual delay to their travel time, relative to the uncontrolled scenario. Clearly, this situation is undesirable.

While the absolute fuel saved is proportional to the distance $d$, the relative fuel savings $R^{\Omega}$ and $R$ are invariant with respect to $d$. The reason is that the size of both the local and global domains in the $(x,t)$-plane is proportional to $d^2$, because both the spatial extent and the duration of the control are linear in $d$.

\begin{figure}
\centering
  \subfloat[$\rho_0=0.66\rho_c=33\>$veh/km]{%
    \includegraphics[width=.48\textwidth]{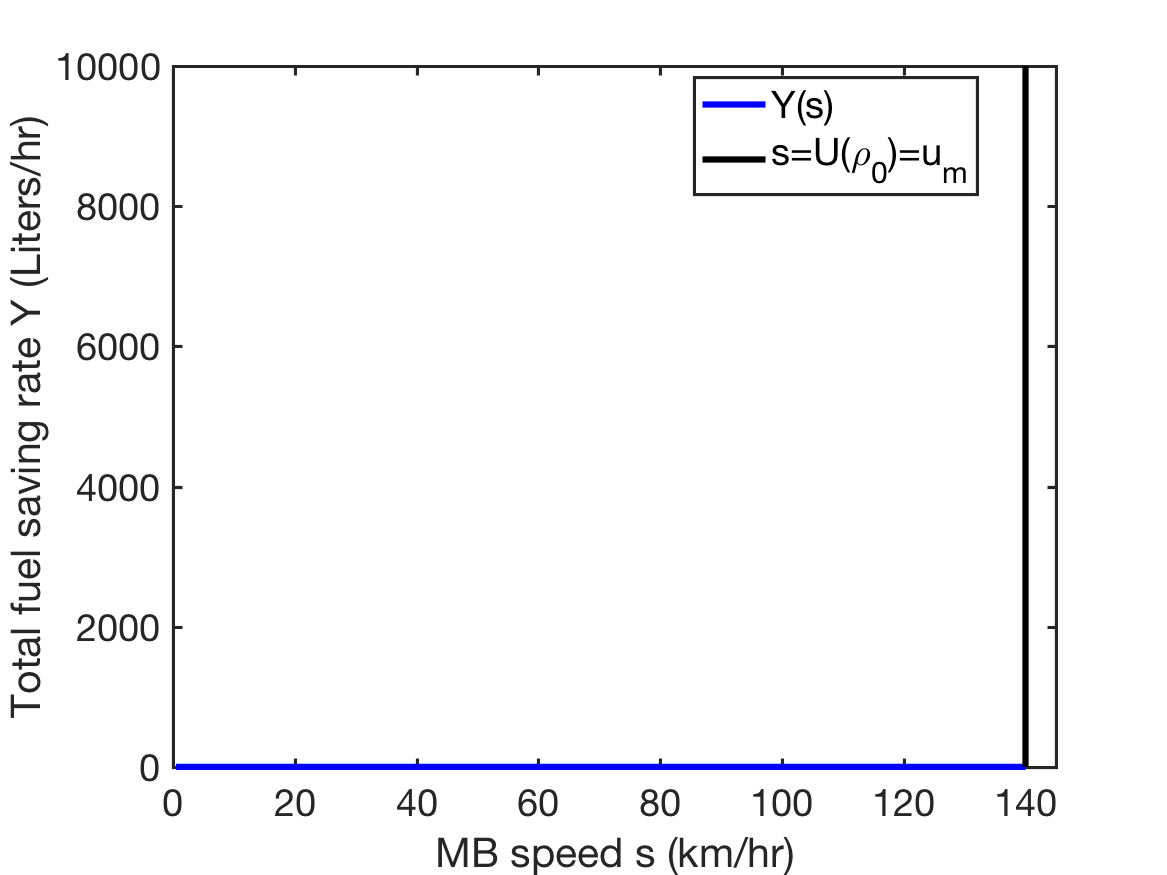}
}\hspace{0in}
  \subfloat[$\rho_0=0.74\rho_c=38\>$veh/km]{%
    \includegraphics[width=.48\textwidth]{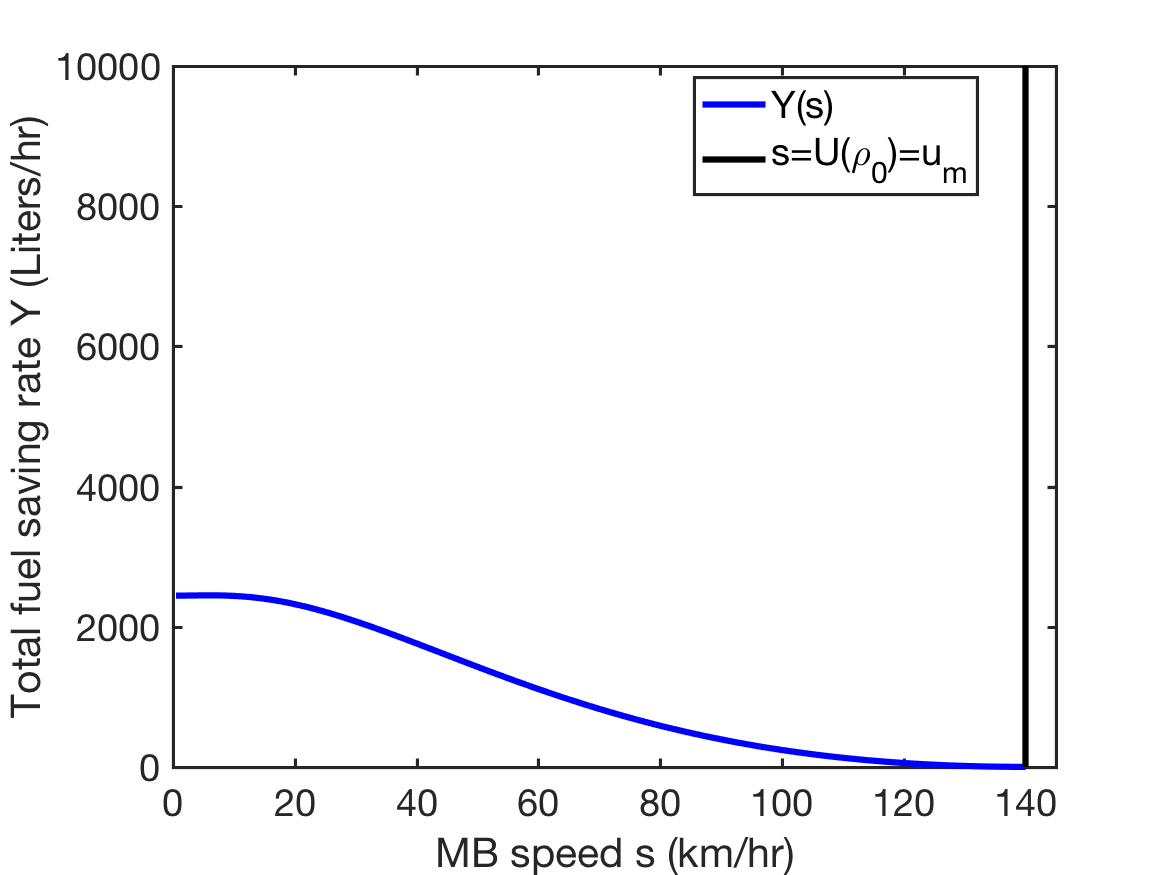}
}\hfill
\\
  \subfloat[$\rho_0=0.86\rho_c=43\>$veh/km]{%
    \includegraphics[width=.48\textwidth]{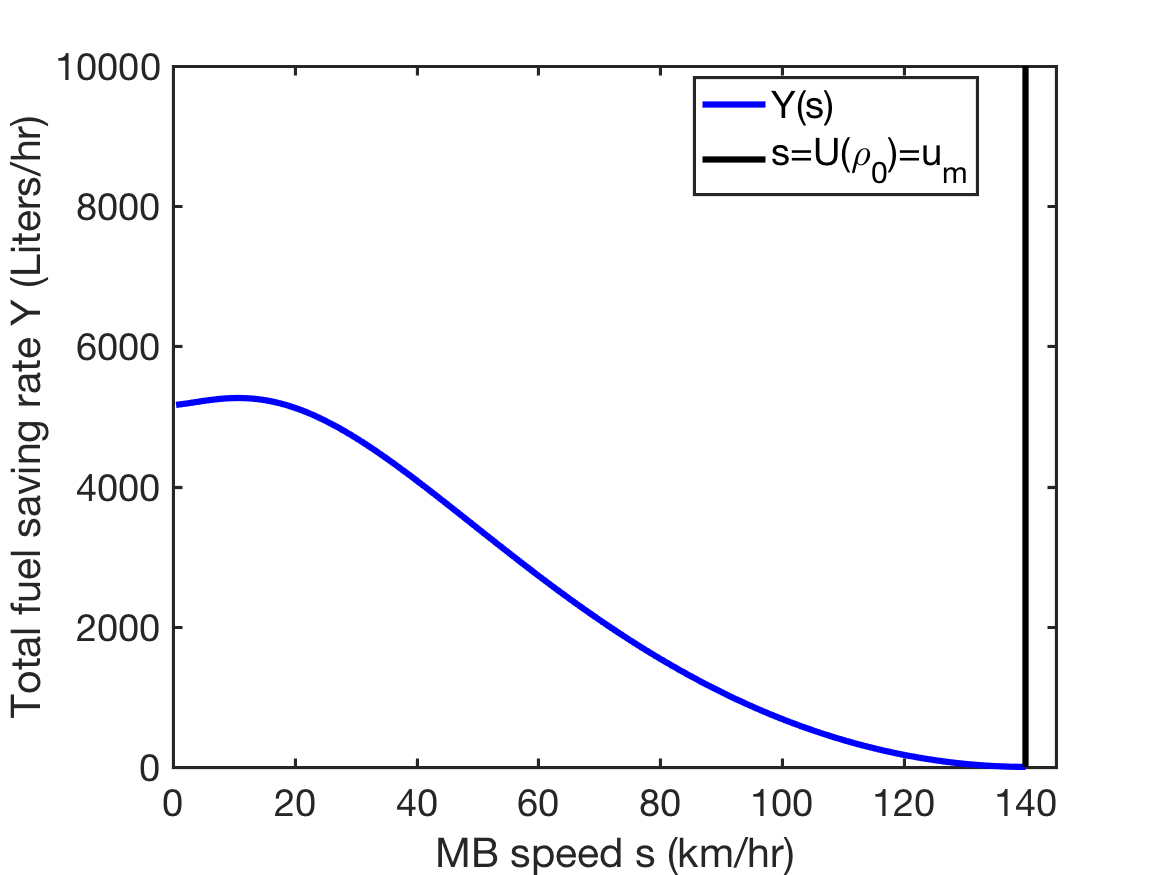}
}\hspace{0in}
  \subfloat[$\rho_0=0.96\rho_c=48\>$veh/km]{%
    \includegraphics[width=.48\textwidth]{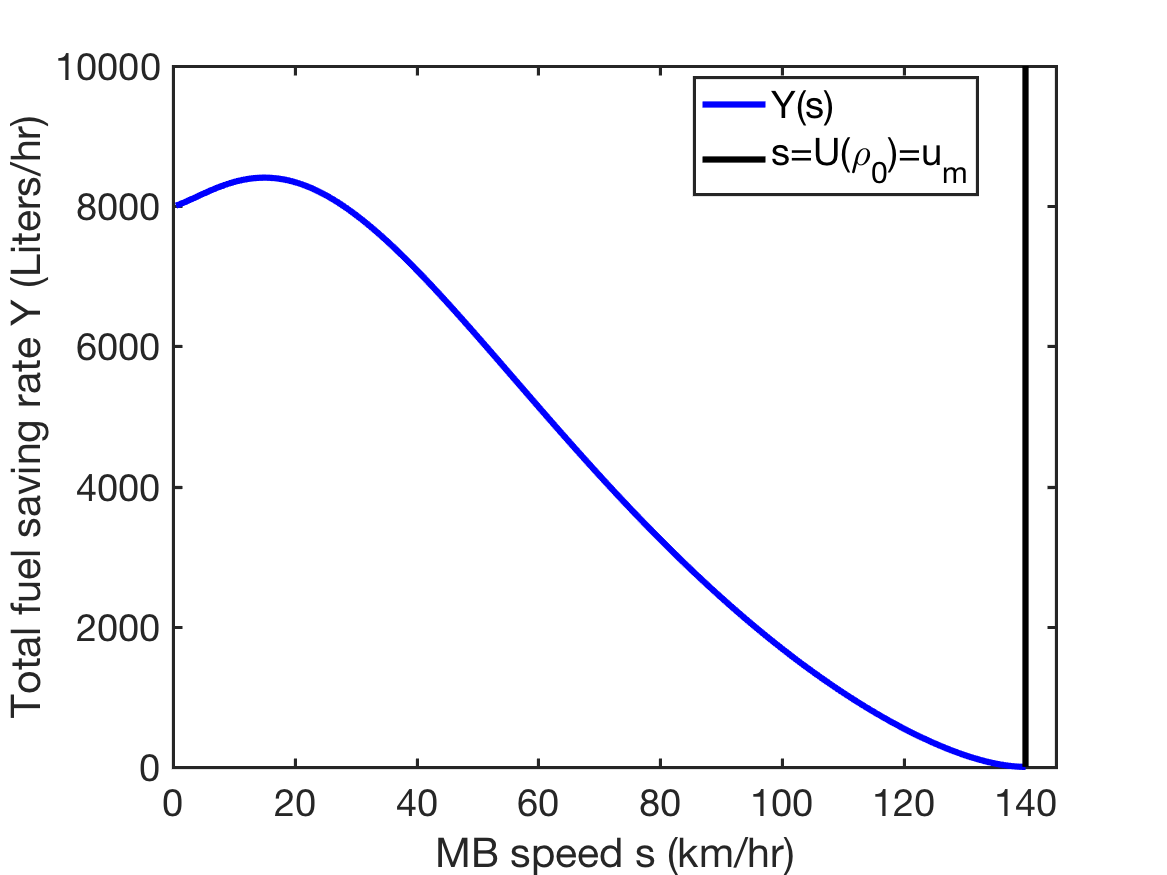}
}\hfill
  \caption{Total fuel saving rate vs.~the speed of the MB, for four different initial traffic densities.}\label{fig:FCS_vs_s_total}
\end{figure}

\begin{figure}
\centering
  \subfloat[$\rho_0=0.66\rho_c=33\>$veh/km]{%
    \includegraphics[width=.48\textwidth]{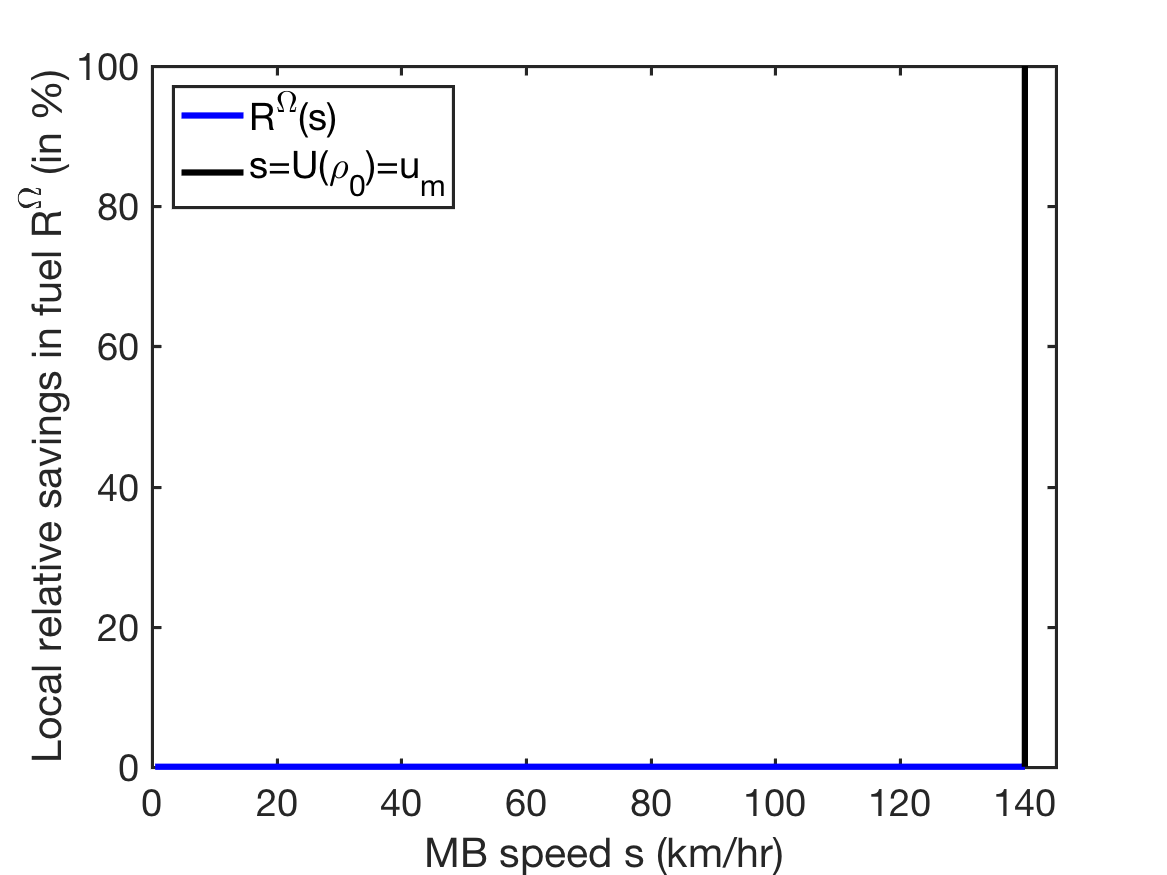}
}\hspace{0in}
  \subfloat[$\rho_0=0.74\rho_c=38\>$veh/km]{%
    \includegraphics[width=.48\textwidth]{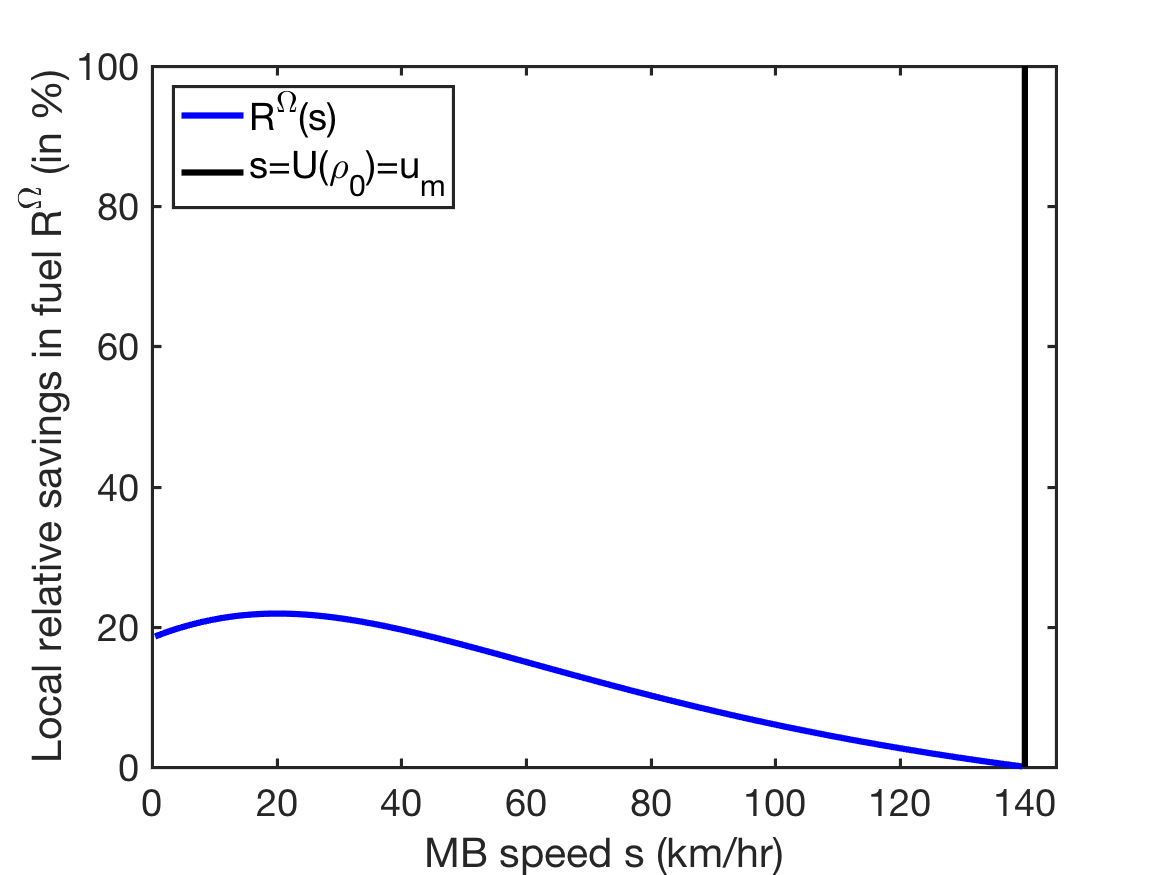}
}\hfill
\\
  \subfloat[$\rho_0=0.86\rho_c=43\>$veh/km]{%
    \includegraphics[width=.48\textwidth]{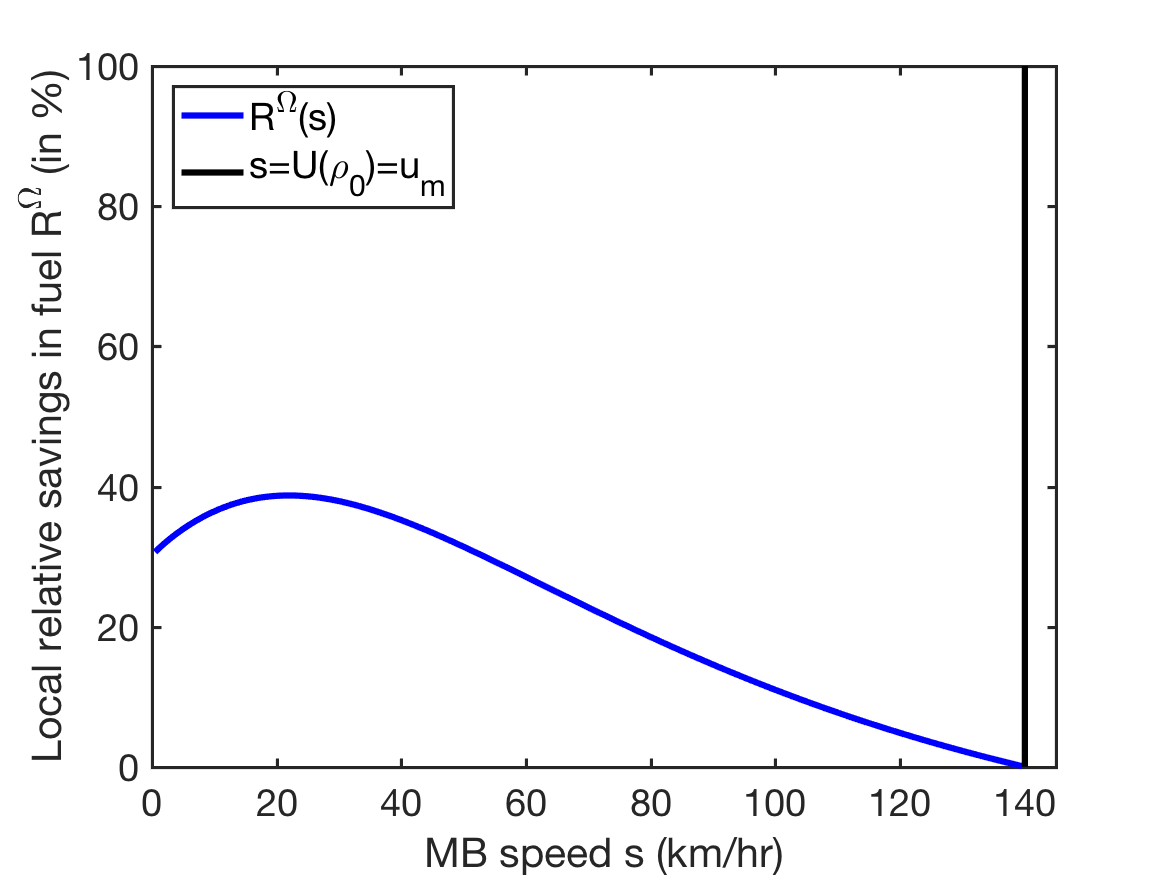}
}\hspace{0in}
  \subfloat[$\rho_0=0.99\rho_c=48\>$veh/km]{%
    \includegraphics[width=.48\textwidth]{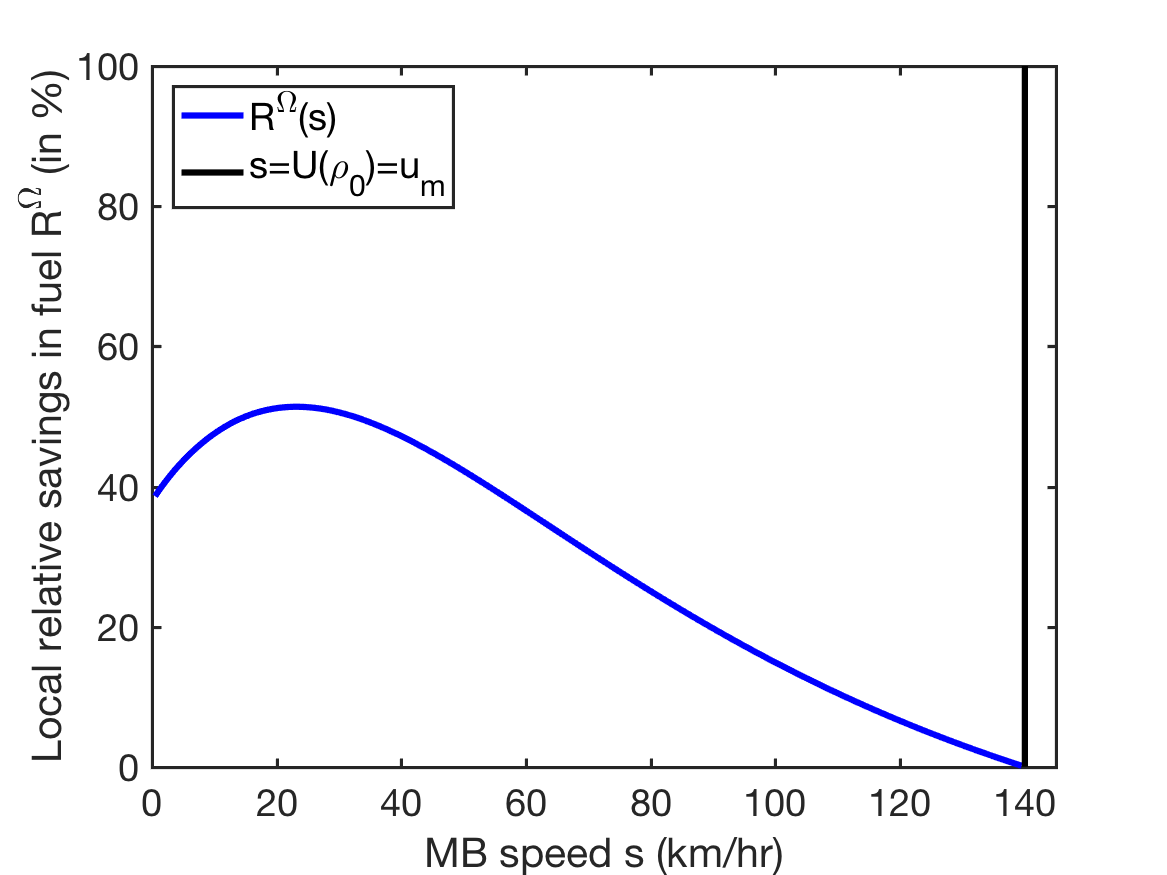}
}\hfill
  \caption{Local relative savings in fuel vs.~the speed of the MB, for four different initial traffic densities.}\label{fig:FCS_vs_s_local}
\end{figure}

\begin{figure}
\centering
  \subfloat[$\rho_0=0.66\rho_c=33\>$veh/km]{%
    \includegraphics[width=.48\textwidth]{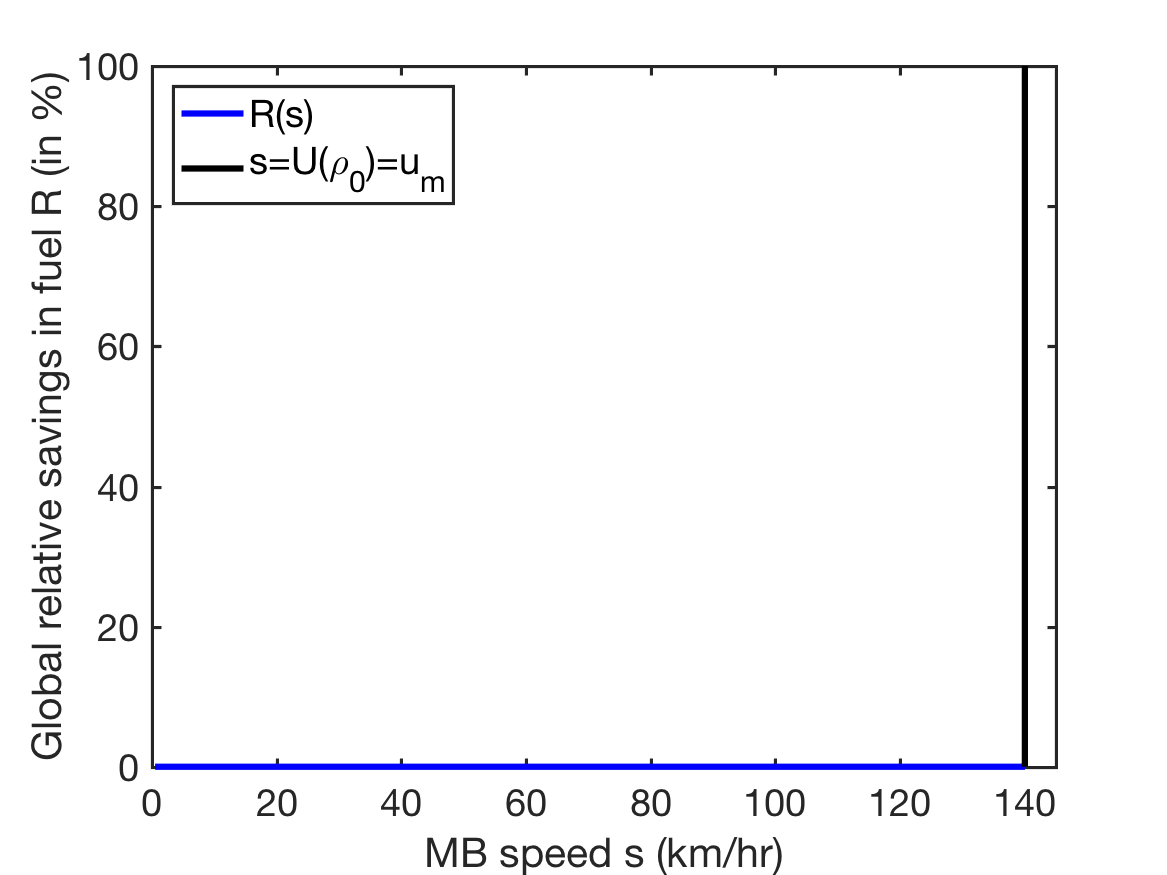}
}\hspace{0in}
  \subfloat[$\rho_0=0.74\rho_c=38\>$veh/km]{%
    \includegraphics[width=.48\textwidth]{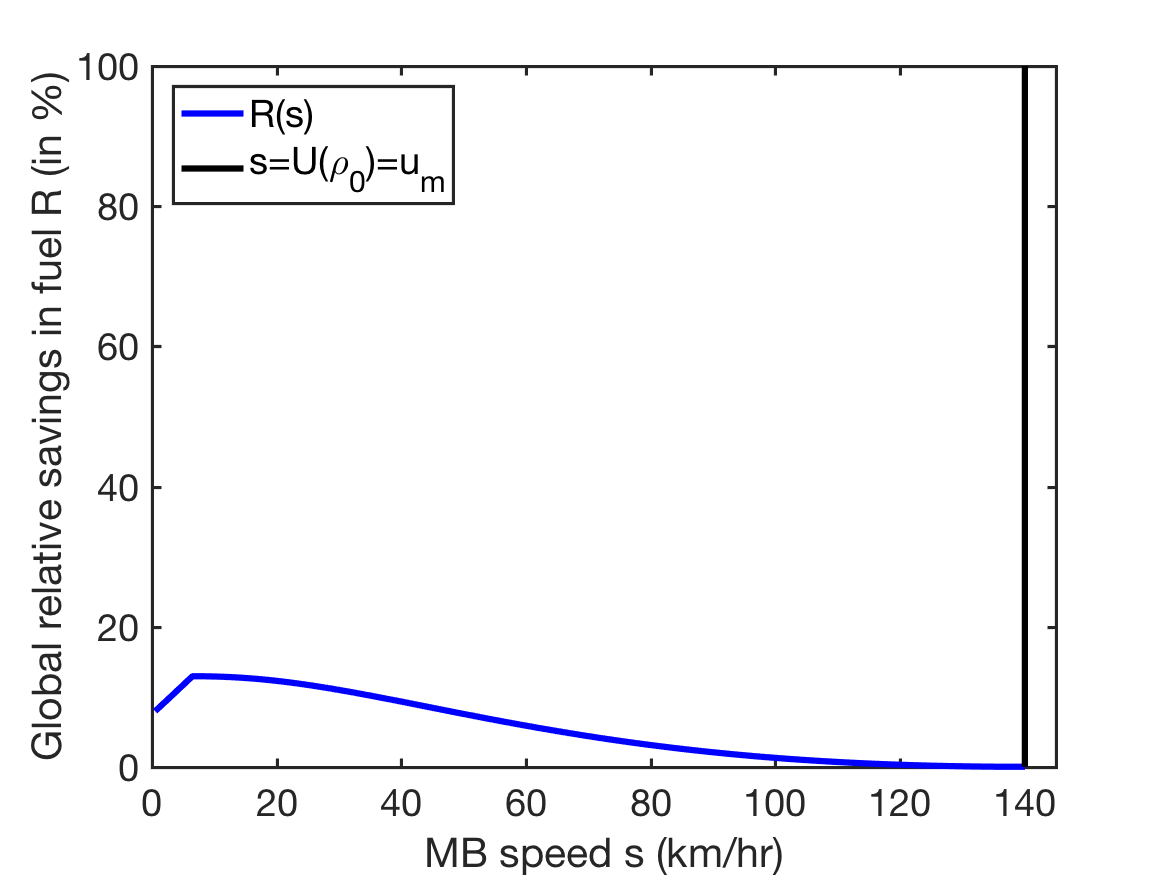}
}\hfill
\\
  \subfloat[$\rho_0=0.86\rho_c=43\>$veh/km]{%
    \includegraphics[width=.48\textwidth]{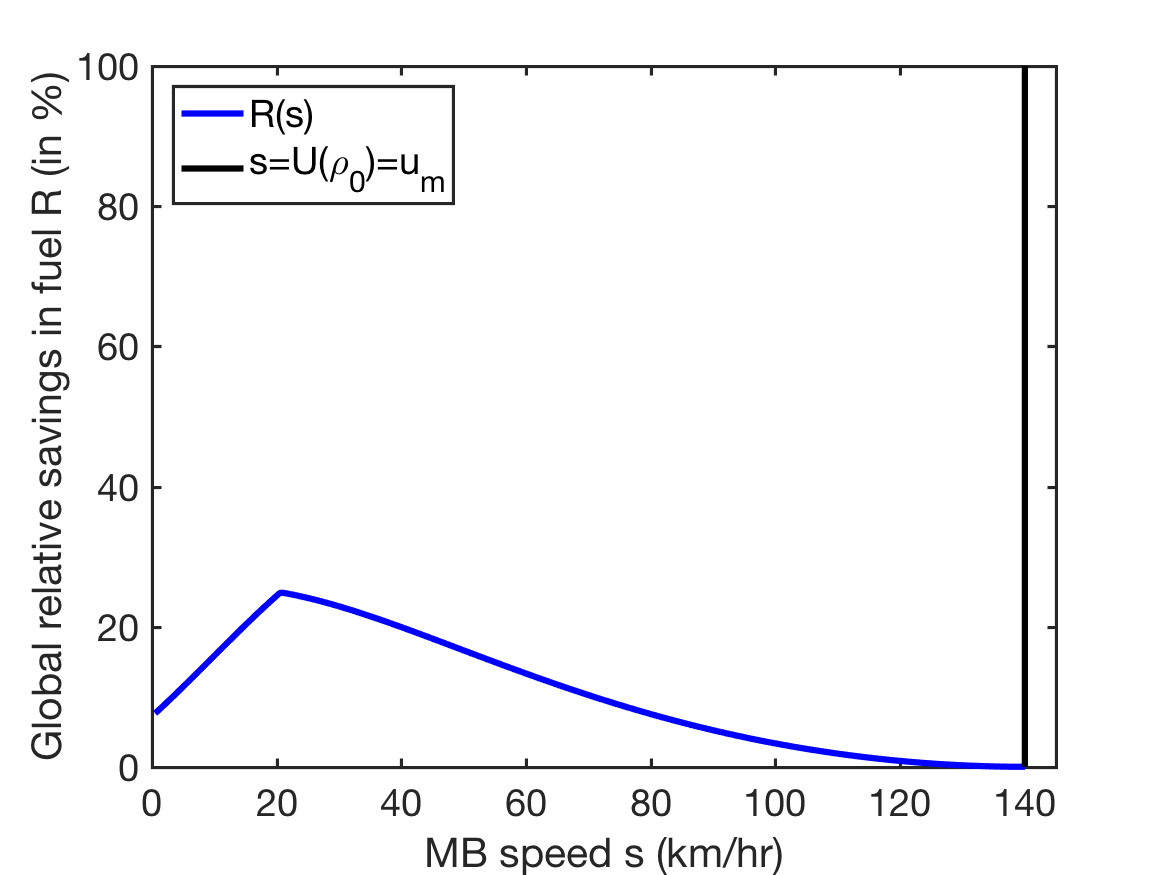}
}\hspace{0in}
  \subfloat[$\rho_0=0.96\rho_c=48\>$veh/km]{%
    \includegraphics[width=.48\textwidth]{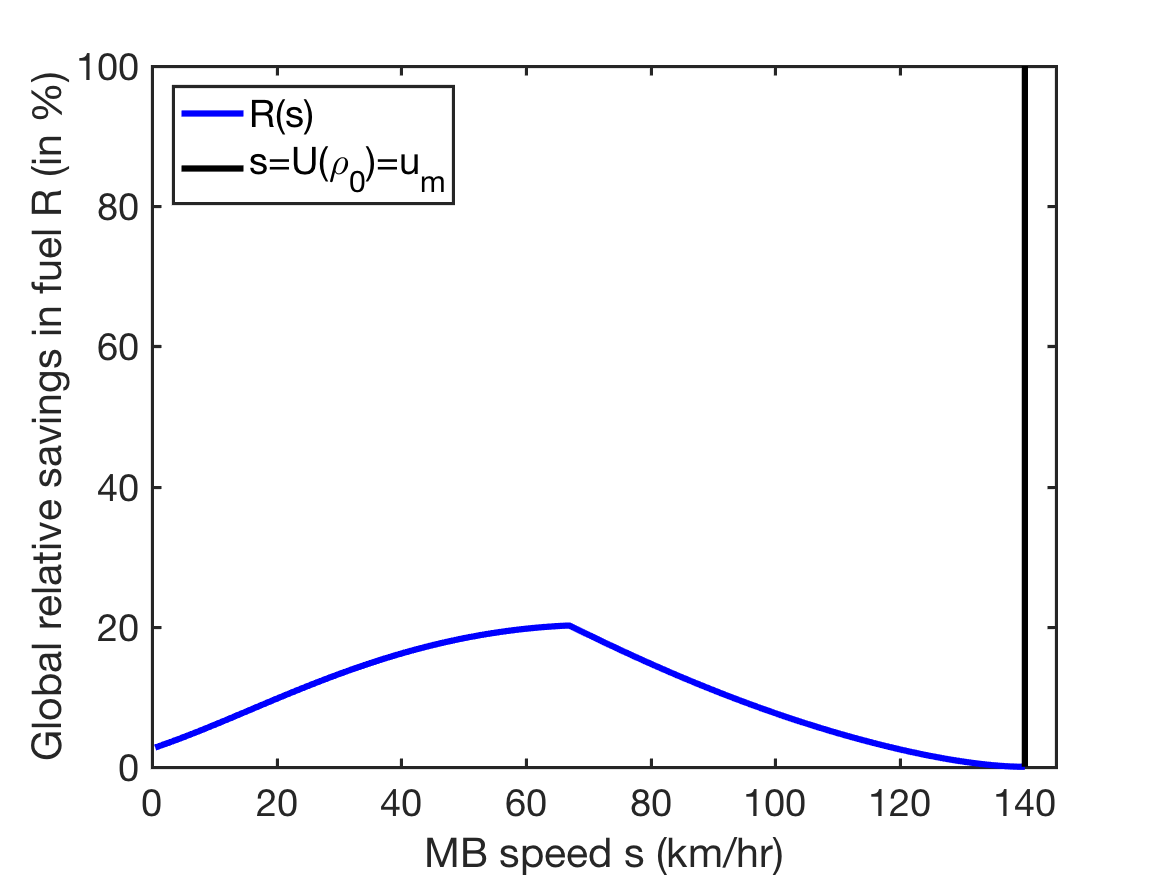}
}\hfill
  \caption{Global relative savings in fuel vs.~the speed of the MB, for four different initial traffic densities.}\label{fig:FCS_vs_s_global}
\end{figure}

%---------------------------------------------------------------------------
\subsection{Effect of the Moving Bottleneck's Speed}
\label{subsec:Optimal_MB_Speed}
%---------------------------------------------------------------------------
The speed $s$ of the MB determines how much it impacts the traffic on the highway. Higher values of $s$ make the effect of the MB milder, whereas low values of $s$ make the MB behave more like a FB, thus inducing drastic velocity changes in its vicinity. Because all quantities $\rho_1$, $\rho_2$, $s_1$, $s_2$, $s_\text{b}$, and $s_\text{low}$ depend on $s$, the fuel consumption measures $Y$, $R$, and $R^{\Omega}$ vary with $s$ as well.

Figure~\ref{fig:FCS_vs_s_total} shows a plot of the fuel savings rate $Y$ vs.~the MB speed $s$, for four different initial densities $\rho_0$ in the free flow regime. In all calculations, the initial distance $d$ between the control vehicle and the highly congested region is 40 km. At a free flow speed of 140 km/hr, this distance is traveled in a little above 15 minutes. In the first case ($\rho_0=33$ veh/km), shown in panel (a), the MB has no effect on the traffic flow for any choice of $s$, because $\rho_0 \leq \beta \rho_c$, i.e., all vehicles can pass the MB without being held back. Therefore $Y = 0$. In the other three cases, we obtain optimal MB speeds $s^*$ for which $Y$ is maximal.

Figures~\ref{fig:FCS_vs_s_local} and~\ref{fig:FCS_vs_s_global} show the corresponding plots for the relative fuel savings, $R^\Omega$ vs.~$s$ and $R$ vs.~$s$, respectively. The same four choices of initial densities $\rho_0$ are considered. The distance $d$ is irrelevant for these two quantities. As in Fig.~\ref{fig:FCS_vs_s_total}, the lowest density renders the MB ineffective, therefore $R^\Omega(s) = 0 = R(s)$ as well. Note that the graphs in Fig.~\ref{fig:FCS_vs_s_global} possess a kink, because the area of the global domain $[z,0] \times [t_1,t_4]$, as a function of $s$, is not differentiable.

As one can see in Figures~\ref{fig:FCS_vs_s_total},~\ref{fig:FCS_vs_s_local}, and~\ref{fig:FCS_vs_s_global}, the maximal FC reduction generally occurs at very low MB speeds. In practice, safety concerns will prevent one from operating a MB at such low speeds. Therefore, the results suggest that for the given situation, within reasonable ranges of MB speeds, the functions $Y(s)$, $R^\Omega(s)$, and $R(s)$ will generally be strictly decreasing. As a consequence, one would generally want to operate the MB at the slowest possible speed that is deemed safe.

For example, if we disallow the controlled vehicle to drive at a speed below $70\%$ of the speed of the ambient equilibrium traffic flow, we obtain a maximum total fuel saved of $Y=1826$ Liters/hr, and maximum relative fuel savings of $R^\Omega=15.82\%$ and $R=8.27\%$, for the initial density $\rho_0=48$ veh/km, and $d=40$ km.

\begin{figure}
\centering
  \subfloat[$\rho_0=30\>$veh/km]{%
    \includegraphics[width=.48\textwidth]{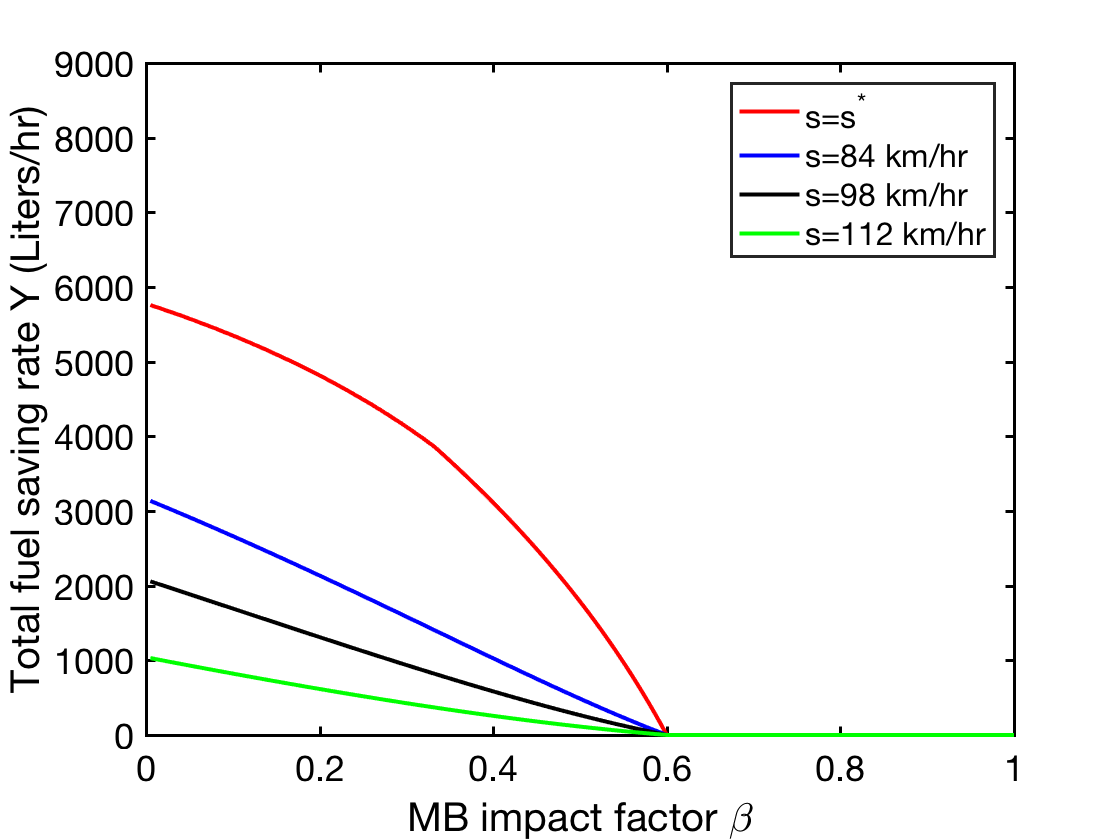}
}\hspace{0in}
  \subfloat[$\rho_0=45\>$veh/km]{%
    \includegraphics[width=.48\textwidth]{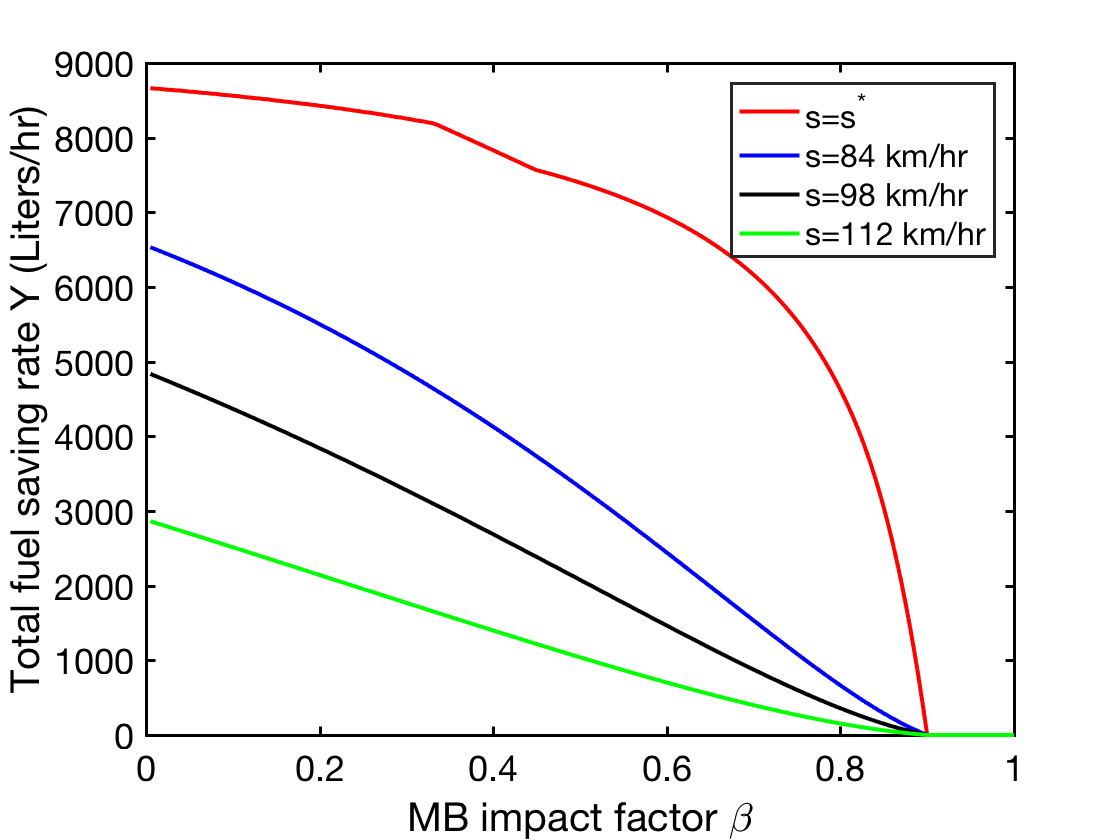}
}\hfill
  \caption{FC rate reduction vs.~the MB impact factor $\beta$.}\label{fig:FCS_vs_beta}
\end{figure}

%---------------------------------------------------------------------------
\subsection{Effect of Parameter $\beta$}
\label{subsec:Effect_of_beta}
%---------------------------------------------------------------------------
In the previous sections, we have assumed the natural value $\beta=\frac{2}{3}$ (2 of the 3 lanes are blocked). In this section, we examine how the results vary if we take a different value for $\beta$. For instance, values of $\beta<\frac{2}{3}$ could arise if lane-changing in the wake of the MB produce extra ``friction'' effects.

Figure~\ref{fig:FCS_vs_beta} shows plots of the fuel savings rate $Y$ vs.~$\beta$ on the interval $0<\beta<1$, for two different densities $\rho_0$, and for four different choices of MB speeds $s$, namely: $s=s^*$ (the optimal MB speed at which $Y$ is maximized; which may be very low), $s=84$ km/hr, $s=98$ km/hr, and $s=112$ km/hr. The MB works better (i.e., provides a larger FC saving) for smaller values of $\beta$. The reason is that the smaller $\beta$, the larger the potential impact of the MB on the traffic state. Note that the red curves corresponding to $s=s^*$ are affected by $\beta$ is two ways: first, $Y$ directly depends on $\beta$; and second, $Y$ depends on $s^*$ and $s^*$ depends on $\beta$. That last dependence $s^*(\beta)$ can be complicated; in fact, for the considered fuel consumption model, it may possess segments where $s^*=0$. This segment is visible in panel~(b) between $\beta=0.33$ and $\beta=0.44$.

The plots in Fig.~\ref{fig:FCS_vs_beta}, panel~(a) show that, while the friction-less choice $\beta=\frac{2}{3}$, considered above, did not lead to any effects due to the MB (for the given low density $\rho_0$), lower values of $\beta$ render the MB effective. Panel~(b) shows that already some mild friction effects could significantly increase the effectiveness of the MB control. For instance, when $\rho_0=45$ veh/km, at $s=112$ km/hr, a reduction of $\beta$ from $\frac{2}{3}$ to $\frac{1}{2}$ roughly doubles the total amount of fuel saved.

%===========================================================================
\vspace{1.5em}
\section{Further Benefits of Moving Bottleneck Control}
\label{sec:Further_Benefits}
%===========================================================================
The investigations above have used the total fuel consumption of the vehicles on the road to quantify the benefit of the traffic control via a MB. The key reason for doing so is that there are precise data available for the vehicles' fuel consumption rate as a function of their velocity (see \S\ref{subsec:Fuel_Consumption}). However, in reality there are greater benefits to the proposed controls than captured by the reduction in consumed fuel. Here, we outline some of them.

The effectiveness of the MB control stems from the fact that for some vehicles, it replaces the rapid transition from high vehicle velocities to very slow traffic flow (uncontrolled case) by two less severe velocity transitions. The inserted middle state (of density $\rho_1$) results in less fuel wasted due to high speed air drag.

In addition, the presence of the middle state also reduces air pollution. A direct reduction in vehicle emissions results from the reduction in consumed fuel. In addition, emissions due to acceleration and deceleration will be reduced as well. The LWR model considered in this paper makes the simplifying assumption that vehicles reach their equilibrium velocities instantaneously and always maintain them precisely. However, in real life vehicles will accelerate and decelerate; and in particular in the highly congested state $\rho_\text{b}$, stop-and-go traffic tends to arise \cite{FlynnKasimovNaveRosalesSeibold2009,SeiboldFlynnKasimovRosales2013}, for which the LWR model captures only the effective average bulk flow properties correctly. The data in \cite{ahn2002estimating} suggests that fuel consumption, and particularly emission rates of HC, CO, and $\text{NO}_\text{x}$ are much higher when vehicles are accelerating, compared to when they are traveling at constant velocities. This implies that the MB, by reducing the length of the traffic jam, will reduce the amount of harmful emissions, that contribute to air pollution.

Moreover, traffic jams (the $\rho_\text{b}$ state) are associated with severely elevated \emph{local} air pollution, because a lot of vehicles are localized to a small area, many of which may be idling \cite{Levy2010}. A quantitative link between traffic congestion, reduced air quality (direct measurements of $\text{PM}_\text{2.5}$, $\text{PM}_\text{10}$, and $\text{NO}_\text{x}$), and health related problems and respiratory symptoms has been provided in \cite{kim2004traffic}.

In the same fashion, further adverse effects of unsteady traffic flow in the high density state $\rho_\text{b}$, such as wear and tear (brakes), or the risk of vehicle collisions, are mitigated by the proposed MB control as well. It can therefore be stated that the benefits of the traffic control strategies developed in this paper go far beyond a mere reduction in fuel consumption.

%===========================================================================
\vspace{1.5em}
\section{Conclusions}
\label{sec:Conclusions}
%===========================================================================
In this paper, we have combined the theory of moving bottlenecks (MBs) and vehicular fuel consumption to suggest a new methodology of traffic control that reduces the overall fuel consumption on highways in certain situations. While the proposed methodology (creating controlled MBs) can in principle be implemented via human-controlled vehicles (e.g., police cars), it carries particular promise in the context of autonomous vehicles (AVs) that will be in the traffic stream in a few years. The proposed ideas work with a single AV, so they are amenable already at extremely low AV penetration rates.

Also, we focused in this paper only on fuel consumption to measure the impact of the MB. We found that to get positive fuel savings, the amount of vehicles traveling at very high speeds must be reduced, where air drag is the main factor for high high fuel consumption. It is important to note that, in the presence of a FB downstream, slowing down fast vehicles will not delay them, but rather only change their speed profile in time. Another positive effect the MB will have in reality is the reduction of air pollution, see the discussion in \S\ref{sec:Further_Benefits}.

Finally, the approach of reducing FC via MBs has been studied quantitatively, using realistic FDs and FC curves. Various FC reduction curves have been provided, suggesting optimal values for the speed of the MB that maximize the reduction in the overall FC. The actual FC reduction achieved by the MB varies strongly with the traffic situation. Considering certain safety constraints, FC reduction rates of more than 8\% can be achieved when considering the full jam produced by a traffic incident. Likewise, when restricting the FC reduction only to the parts of the jam that can be affected by the MB, a FC reduction of about 16\% is possible. Also, considering a realistic distance $d$, we were able to get fuel savings of about $1800$ Liters/hr. Due to the nature of the traffic control, the reductions in FC come at very small cost.

%===========================================================================
\vspace{1.5em}
\bibliographystyle{plain}
\bibliography{references_complete}
%===========================================================================

\end{document}